# Scientific Preparations for Lunar Exploration with the European Lunar Lander

J.D. Carpenter[*1], R. Fisackerly, D. De Rosa, B. Houdou

[*]ESA ESTEC, Keplerlaan 1, 2401 AZ, Noordwijk, The Netherlands; [1]e-mail james.carpenter@esa.int

**Abstract**

Recent Lunar missions and new scientific results in multiple disciplines have shown that working and operating in the complex lunar environment and exploiting the Moon as a platform for scientific research and further exploration poses major challenges. Underlying these challenges are fundamental scientific unknowns regarding the Moon's surface, its environment, the effects of this environment and the availability of potential resources. The European Lunar Lander is a mission proposed by the European Space Agency to prepare for future exploration. The mission provides an opportunity to address some of these key unknowns and provide information of importance for future exploration activities.

Areas of particular interest for investigation on the Lunar Lander include the integrated plasma, dust, charge and radiation environment and its effects, the properties of lunar dust and its physical effects on systems and physiological effects on humans, the availability, distribution and potential application of in situ resources for future exploration. A model payload has then been derived, taking these objectives to account and considering potential payloads proposed through a request for information, and the mission's boundary conditions. While exploration preparation has driven the definition there is a significant synergy with investigations associated with fundamental scientific questions.

This paper discusses the scientific objectives for the ESA Lunar Lander Mission, which emphasise human exploration preparatory science and introduces the model scientific payload considered as part of the on-going mission studies, in advance of a formal instrument selection.

**Keywords**

Moon; south pole; Lunar Lander; exploration; dust; plasma; toxicity; resources; volatiles; radiation

# 1 Introduction

The Moon is an important target location for future exploration by both human and robotic missions. This paper describes the European Lunar Lander mission under development by the European Space Agency (ESA) and the role it can play in enabling scientific research that can support this future exploration.

In the decades following the Apollo and Luna exploration era of the 1960s and 1970s the Moon fell somewhat out of the international spotlight but a multitude of recent missions from the USA, Europe, Japan, China and India have reignited interest in our nearest celestial neighbour. The Apollo programme has been and continues to be a foundation for planetary science. The scientific results obtained through the Apollo programme are great enough in magnitude to eclipse those of every other planetary mission to date and are a cornerstone for our understanding of the Moon (Crawford et al. 2012; this issue; Jaumann et al, 2012; this issue). Recent missions however have shown that our understanding of the Moon is far from complete, the Moon is far more complex than we imagined and much remains to be leaned.

While the fundamental scientific results obtained by these recent missions are significant it is also important to note that their objectives have not been entirely motivated by fundamental science. Many of these missions have been developed in order to prepare for more comprehensive exploration in the future. A good example of this is NASA's Lunar Reconnaissance Orbiter mission (Chin et al., 2007), which was defined with the primary purpose of characterising the Moon, its surface and environment in a manner that could be enabling for future missions. The Chinese Chang é 1 and 2 missions (e.g. Huixian et al., 2005) have also been defined as part of a rapid extension of exploration capabilities in China, which should lead to eventual human missions.

The future plans and goals of multiple agencies and actors in the private sector strongly indicate that Moon exploration is entering a new era. A framework for cooperation in exploration between 14 space agencies entitled The Global Exploration Strategy was published in 2007 (The global exploration strategy, the framework for cooperation, 2007). Arising from this was the International Space Exploration Coordination Group (ISECG), which in 2011 published its Global Exploration Roadmap (ISECG, 2011). This roadmap clearly shows the Moon as an essential stepping stone for exploration by both humans and robots.

Exploration of the Moon offers multiple benefits to its participants. As a target for exploration it is an important stepping stone where technologies can be proven and capabilities can be sought on the way to more distant destination such as Mars, (Cockell, 2010; Crawford, 2004; Mendell and Heydorn, 2004). The Moon also provides a platform where both physiological and psychological challenges of exploration can be addressed (Durante, 2012; this issue; Gosswami et al, 2012; this issue). The Moon is of course itself an important location for performing scientific research and can be an important location for addressing multiple scientific goals (e.g. Committee on the Scientific Context for Exploration of the Moon, 2007; Crawford et al., 2012). The coming era of lunar exploration is likely to yield very significant scientific returns in multiple scientific areas including planetary sciences and geology, space science and plasma physics, fundamental physics, astrobiology, physiology, astronomy and others . It is within the context of preparing Europe to participate in and receive the maximum benefit from future lunar exploration that ESA had defined a Lunar Lander mission, as a precursor mission to future exploration.

## 2  The European Lunar Lander Mission

The primary objective of the Lunar Lander mission is to demonstrate soft, safe, precision landing. This has been identified as an important capability for future exploration in the international context and it is the requirement to demonstrate this capability that drives the mission design (e.g. Fisackerly et al, 2011). Once on the surface however the mission has a second objective, to provide data sets that can be used to help address important unknowns for exploration by performing scientific investigations. The scientific investigations to be performed are defined by the scientific needs of exploration to enable sustainable exploration programmes in the future.

The mission aims to land within a predefined area, several hundred meters in diameter in the South Polar region of the Moon. In this region locations can be identified at which the low sun inclination, coupled with topographic highs lead to sustained periods of solar illumination (de Rosa et al, 2012; this issue). These locations present an opportunity to enable surface mission durations of much longer than one lunar day (15 Earth days), whilst using solar power and batteries as sources of power. Targeting such landing sites allows a sustained period of operations for the Lunar Lander Mission and also drives the technology development for precision landing. Extensive work has been carried out to characterise the locations and sizes of these landing sites and to determine the extent of surface hazards that may be encountered at the surface there (de Rosa, 2012; this issue).

A key characteristic of these landing sites will be short periods of darkness, experienced when the Sun is obscured by topographic high points on the local horizon. During these short periods of darkness temperatures are likely to drop to as low as 100K for periods of hours or tens of hours (de Rosa, 2012). Surviving these periods poses technological challenges for the thermal and power subsystems of the lander and drives the development of a highly optimised and efficient spacecraft.

During surface operations the mission will perform complex tasks, including scientific investigations and robotic manipulation of payloads and lunar regolith. These operations too will drive new technologies and capabilities, which can be a precursor to those required by future exploration activities.

The potential objectives and priorities for exploration enabling science to be carried out by the mission were identified through consultation with a range of scientific communities. The results of this consultation were reported in an ESA document (Lunar Exploration Definition Team, 2009) and describe scientific objectives and requirements, for exploration precursor missions to the Moon, in areas as diverse as radiation biology, dust toxicology, dust properties and effects, astronomy and lunar geology and geophysics. Areas for consideration are also described below and more comprehensively by parallel papers published in this issue (e.g. Anand et al., Bergmann, 2012; Crawford et al, 2012; De Vos et al., 2012; Durante, 2012; Klein-Wolt, et al. 2012; ; Reitz et al., 2012; Sheridan et al., 2012; Staufer et al., 2012; Linnarson et al., 2012; Zarka et al., 2012).

Scientific investigations considered in the frame of preparing for future exploration included those in terrestrial laboratories, those in analogue environments (e.g. terrestrial analogue sites, international Space Station, Parabolic Flights etc.) and those in situ on precursor missions like the Lunar Lander. Potential experimental techniques to be applied in situ were then defined and the requirements for a system to implement these techniques in situ were defined. These were then compared with the expected capabilities of the mission to define a likely suite of instruments to address high priority objectives, which may be feasibly addressed by the Lunar Lander mission and a potential suite of instruments to be considered for the mission. This model payload, which represents a working assumption of the scientific payload of the mission, is described below and has been applied during the study phase of the mission in advance of a formal instrument selection. The boundary conditions considered for the mission were identified through a series of five Phase A mission study activities performed between 2008 and 2010 under the title of MoonNEXT (e.g. Carpenter et al. 2008). The boundary conditions for a Lunar Lander mission derived from these studies are described below. These conditions have been applied in the subsequent mission design phase currently underway.

## 2.1 Mission Outline

The Lunar Lander space segment is comprised of a single spacecraft launched from Kourou on-board a Soyuz launch vehicle not later than 2018. It is injected onto a highly elliptical intermediate orbit by the launcher's Fregat-MT upper stage. The Lander then uses its own chemical propulsion system to complete the transfer to the Moon. After injecting itself into a polar orbit around the Moon, the Lander proceeds to acquire the 100km by 100km low circular orbit in which it will wait before beginning the landing sequence.

The low Lunar orbit phase is used to provide sufficient time for the correct phasing and orientation of the Sun, Earth and Moon for landing, while accommodating mission constraints from launch and transfer, as well as to execute any check out procedures prior to descent. While in low Lunar orbit, ground tracking is used to establish an accurate determination of the Lander's position and velocity, which are uploaded to the spacecraft prior to the initiation of the descent and landing phase and are crucial in order to ensure the Lander's knowledge of its state has the greatest possible accuracy.

Once Earth, Sun and Moon are properly aligned the Lander shall begin the landing sequence and put into action the integrated suite of sensors, image processing, guidance, navigation and control algorithms which form the heart of the descent and landing system. The decent and landing sequence, described below is illustrated in Figure 1.

Firing its engines once over the Moon's North Pole, the lander executes descent orbit initiation and begins coasting for one final half orbit towards the South polar region. During this coasting phase the Lander uses Optical Absolute Navigation (OAN).  OAN uses the onboard computer to process images of the surface taken by a dedicated navigation camera and extract remarkable landmarks, which can be natural features like craters or specific image features like edges or corners. These landmarks are then identified, by matching them with a set of landmarks stored on board in a database that is built in advance of the mission using data from lunar orbiters, such as Lunar Reconnaissance Orbiter (LRO) or Kaguya. The landmark database referenced by the system  includes the position of the observed landmarks to an accuracy of tens of metres. With this information the , the syste is able to significantly improve its knowledge of the lander's position.

As the Lander begins to close on the South Polar Region it initiates the main braking phase by firing all of its main engines and exerting maximum thrust. During this period the Lander uses Visual Relative

Navigation (VRN) to extract velocity information by tracking features in images taken by an on board camera. This velocity information, combined with other navigation sensors, ensures that the Lander maintains an accurate knowledge of its position with respect to the landing site.

On approach to the landing site the Lander modulates the level of thrust exerted by the engines, progressively shutting down the main engines and compensating with its assist engines operating in a pulsed mode. The Lander is thus able to precisely control its descent in a fuel efficient way and to keep itself on track for an accurate landing.

Once the landing site is in view of the Lander, the surface is scanned by the on board sensors for potential hazards including slopes, boulders and shadows. If the nominal landing site is identified as hazardous the Lander has can autonomously retarget to a new, safe landing site. The terminal phase sees the Lander execute a vertical descent towards the safe landing site, and is completed with touchdown on the Lander's four landing legs.

The descent and landing phase of the mission requires that critical decisions be taken in real time and thus a high degree of on board autonomy is required. These autonomous systems select the safe landing site, while continuously monitoring the spacecraft status and available resources. In line with the Beagle-2 Inquiry Report (Bonnefoy et al., 2004), status information will be transmitted directly to Earth during the most critical phases of descent and landing via a Direct to Earth link.

Once successfully on the Lunar surface the Lander begins operations with the deployment of the most important surface elements such as the camera mast and the main antenna. The Lander stores the complete information associated with the descent and landing on board and transmits this full data set back to Earth immediately, once nominal communications are established. In doing so the primary technology demonstration objective can be confirmed. Alsoof high priority is reconnaissance of the local terrain, specifically the horizon, by the surface cameras to identify features which may obscure the Sun at some point in the mission timeline.

Communications are established through a direct-to-Earth link, which is not dependent on any relay spacecraft, but which must account for the 28-day cyclic rising and setting of the Earth over the local horizon.

The surface mission phase continues with the deployment of the relevant payloads, in some cases relying on the robotic arm. One element under consideration is a Mobile Payload Experiment, a small robotics demonstration payload , which is deployed onto the surface and begins its operations supported for communications by the Lander. Sampling activities are commanded from Earth during communications windows, during which the samples are analysed by the associated instruments. Other payloads investigating the surface characteristics store data on board while the Earth is not visible; this is then transmitted back to ground when the Earth rises again above the horizon.

Depending on the landing site and the local topography, the Lander will experience periods of darkness. Short darkness periods caused by specific surface features such as boulders or distant peaks will require the Lander to enter into a survival mode in which operations are reduced such that energy output is at a minimum and the on board battery's charge conserved as much as possible (de Rosa et al., 2012; this issue). Through surviving short darkness periods the Lander can support longer overall surface operations. The nominal mission lifetime on the Lunar surface is expected to be 4-6 months though the final duration of the mission is dependent on the specific properties of the precise landing point reached.

## 2.2 Lunar Lander Spacecraft

The Lunar Lander spacecraft configuration is shown in Figure 2. It is made up of a cylindrical body accommodating solar panels around its circumference and enclosing the propellant tanks. At the core of the Lander is a Central Avionics Box which houses and thermally insulates critical equipment which must survive on the surface during any darkness periods. The base of the Lander is made up of the thruster platform which accommodates the five main and six assist engines.

Landing legs, folded at launch, are deployed and latched around the Lander body. The landing legs ensure the platform stability at landing on slopes of up to 15° and creates clearance under the thrusters compatible with landing on rocks of up to 50cm in height. Hazards exceeding these limits shall be avoided by the hazard detection and avoidance system during the approach phase. The HDA system uses the navigation camera images in order to detect shadowed areas on the surface and areas of high texture, which correspond to rough terrain. The system also uses a model of the terrain generated on-board by a scanning Light Detection And Ranging (LIDAR) instrument, in order to detect steep slopes and rough terrain features (e.g. boulders). If the surface area towards which the lander is heading following its nominal trajectory is found to be unsafe by the HDA system, the system commands a manoeuvre to new, safe area, taking also into account the propellant left and the maximum attitude rates that can be sustained. Once above the selected landing site, the Lander executes a vertical descent towards the surface, until touchdown on the Lander's four landing legs.

The top platform hosts several elements including the thermal radiators, payloads, attitude thrusters, antenna and camera mast. Navigation and hazard detection sensors are accommodated around the Lander body such that they have an un-obscured view of the surface. A robotic arm can extend its reach from the top platform down to the surface to deploy payloads and to acquire regolith samples.

# 3   Scientific Preparations for Lunar Exploration

Lunar explorers of the future face operations and habitation for sustained periods of time in a complex environment whose properties and effects on both systems and people are currently poorly constrained. Below we summarise some of the major unknowns, for which measurements from the Lunar Lander have the potential to provide vital data points. Readers are directed to specialist papers within this issue for more comprehensive descriptions of the scientific areas in question (e.g. in this issue, Anand et al., Bergmann, 2012; Crawford et al, 2012; De Vos et al., 2012; Durante, 2012; Klein-Wolt, et al. 2012; 2012; Reitz et al., 1012; Sheridan et al., 2012; Staufer et al., 2012; Linnarson et al., 2012; Zarka et al., 2012).

## 3.1   Landing site Characterisation

Future exploration activities may take human explorers away from the well characterised and fairly benign landscapes of the Apollo missions. Landing sites at locations such as the polar regions and South Pole-Aitken Basin, or the sites on the Lunar farside are perhaps of the greatest intrinsic scientific interest (e.g. Jaumann et al., 2012; this issue) but present very significant challenges for landing and surface operation. Characterisation of the topography, rock and crater distributions and other characteristics of these landing sites, beyond what can be accomplished from orbit, and in order to validate the interpretation of orbital data sets, are of great importance for the design and planning of future missions and the associated infrastructure (de Rosa et al., 2012). To this end one of the objectives of the mission is to characterise the suitability of a potential landing sites for future exploration by characterising one such site in detail and providing ground truth to inform future interpretation of orbital data on other landing sites.

Characterisation of a southern polar landing site from the surface will require the measurement of several important parameters and the returning of data which allows the revision and validation of models of key surface characteristics. Such parameters and characteristics which should be measured or addressed include illumination environment at the surface including the duration and cycle of light and dark; thermal environment at the surface including local temperatures throughout the illumination cycle including the effects of shadowing by local objects and more distant topography and the influence of the regolith; the distribution of local surface features such as boulders, craters and slopes; and the mechanical properties of the regolith

### 3.2 The dust and plasma environment and its effects

During the brief periods of time spent on the Moon during Apollo it became apparent that lunar dust could pose significant problems for the operations of both people and equipment. Dust was found to adhere to clothing and equipment, it reduced visibility during landings, mechanical devices were severely compromised by lunar dust contamination, optical components were covered with visible dust layers and Apollo astronaut spacesuits became coated with fine-grained dust (Wagner, 2006). Once inside the spacecraft dust caused physiological effects and inhibited vision (e.g. Wagner, 2006, Stubbs et al., 2007a). In addition dust was also found to prevent effective sealing of pressurised and depressurised containers. None of the containers containing rock samples from any of the Apollo missions was found to be able to hold vacuum after return to Earth due, it is expected, to dust grains inhibiting the knife edge indium seals (Taylor et al., 2005). Pressurisation of lunar modules after the initial opening required more oxygen in order to counter the effects of dust on the seals. These dust particles were also found to have an abrasive effect on materials of both spacesuits and mechanical surfaces found in equipment such as rovers. Abrasive effects were also observed on both aluminium and painted surfaces retrieved from the Surveyor 3 lander during the Apollo 12 mission. The abrasion and dust adhesion were related, for the most part, to particles propelled during the landing of the lunar module 183m away (Caroll and Blair, 1972), although a recent reanalysis of surfaces indicates that a fine layer of lunar regolith had coated the materials and was subsequently removed by the Apollo 12 Lunar Module landing rocket (Immer et al., 2011).

The adhesion of dust to both spacesuits and machinery during operations was found to occur by both mechanical and electrostatic means. Mechanical adhesion occurred as a result of the irregular and angular structures of much of the lunar dust and electrostatic adhesion occurred as a result of electrostatic charging. Although adhesion forces between dust particles are primarily the result of van der Waals forces (Katzan and Edwards, 1991), the charging of lunar dust drives the transport and mobility of dust particles and is also important for adhesion. Lunar dust was observed to adhere to painted surfaces with a strength of ~ 104 dynes / cm$^2$ and to metallic surfaces with a strength of ~ $2x10^3$ – $3x10^3$ dynes / cm$^2$ (McKay et al., 1991). Additional measurements of the effects of lunar dust have been provided by recent re-analysis of data from the Dust Detector Experiments (DDEs) deployed during the Apollo 11 and 12 missions (O'Brien, 2008). These data have shown that the Apollo 11 seismometer may have failed following contamination by dust during the ascent of the Apollo 11 module. In addition it was shown that the strength of electrostatic adhesive forces, binding dust to both horizontal and vertical and smooth surfaces, increases as a function of the incident angle of solar radiation indicating that electrostatic effects are important. The transition point, at which electrostatic forces are just sufficient to counter the downward force due to lunar gravity occurs when incident Solar brightness is ~100mW cm$^{-2}$. This corresponds to 70% of full solar intensity at an incident angle of 45°to a given surface. Thus it is suggested that clinging dust experience by Apollo astronauts, at solar elevations much less that 45°, is most likely to have been driven by forces other than electrostatic ones..

An additional effect of dust, little appreciated during the Apollo era, is its toxic potential for humans. An extensive review of the potential toxic effects of lunar dust and the needs for future research is presented by by Linnarson et al. (2012; this issue). Exposure to mineral dust, of various types, has been demonstrated to lead to various pulmonary diseases including cancer and silicosis. The effects of nanometer scale dust particles, which may penetrate deep into the body, may be far wider reaching. The properties of lunar dust, as they are understood, indicate a high likelihood of significant toxic effects. However the combination of the unusual properties of lunar dust, as compared to terrestrial analogues, the uniqueness of the environment in which it resides and the likelihood that existing samples of lunar dust are not fully representative of particles in situ leave major uncertainties. Not least is the surface chemical reactivity of grains, which may result from mechanical fractionation, radiation exposure and solar wind reduction (Loftus et al, 2010). In situ measurements are required to validate terrestrial models for lunar dust toxicity. A further objective for the Lunar Lander then is to make in situ measurements which allow the optimisation and validation of terrestrial investigations into the toxicity of lunar dust.

While the mechanisms for the mobilisation of dust during Apollo were associated with human, lander or machine related activities there is also evidence for a population of particles, whose motion is spontaneous and related to the complex interaction between solar radiation, the dusty lunar surface and both solar wind and magnetospheric plasmas (e.g. Grün et al., 2011, Halekas et al., 2011, Pines et al., 2011). The existence of such a population has been inferred from a number of observations from spacecraft, astronaut observers and experiments. The Surveyor 5, 6 and 7 spacecraft observed a "horizon glow" along the western horizon of the moon following sunset (e.g. Glenar et al., 2011). This glow was attributed to the forward scattering of sunlight by electrostatically levitated dust grains with diameters of <10 µm, at altitudes of 10 – 30 cm. Apollo 17 astronauts reported observing seeing bright "streamers", whose brightness changed on timescales of the order of seconds to minutes, and which extended to altitudes in excess of 100 km. These streamers have been attributed to the sporadic elevation of 100 nm scale particles by an effect referred to as the "dynamic dust fountain" (Stubbs et al., 2007b).

The only direct measurement of dust levitation was performed by the Lunar Ejecta And Meteorites (LEAM) experiment deployed during the Apollo 17 mission (Berg et al., 1976). This experiment was intended to investigate hypervelocity impacts by meteoroids but serendipitously detected impacts by slower (< 1km / s) charged dust particles. The experiment included three sensor systems, facing in the east, west, and vertical directions. Data showed an increase by a factor of 100 in the impact rate for these slow particles on the east and vertical pointing sensors at dawn. The increase began between 3 and 40 hours before the terminator and fell to ~ 0 within 30 hours of the passage of the terminator. While such measurements are a strong indication of dust motion the validity of such a conclusion has been questioned and instrumental effects have been proposed as a mechanism for generating such effects (O'Brien, 2011).

An additional indication of the long term effects of a natural mobile dust population arises from measurements made using the laser retroreflectors placed on the lunar surface during the Apollo missions (Murphy et al., 2010). It has been found that the intensity of reflected laser light from these reflectors has decreased by around an order of magnitude since their emplacement some 40 years ago. The most likely process underlying this loss of performance has been identified as being the gradual deposition of dust.

The nature of physical mechanisms that can drive the levitation and mobility of dust particles across the lunar surface are at present poorly constrained, but must be the result of interactions between these dust particles, solar radiation, "cold" solar wind plasma and "hot" plasma in the Earth's magnetosphere. Dust on the day side of the Moon is exposed to solar Ultra-Violet and X-ray photons and solar wind plasma. Ionisation occurs, dominated by photoionisation. The electrical conductivity of the regolith is typically $10^{-12} - 10^{-13}$ $\Omega^{-1}m^{-1}$ (McKay et al 1991), sufficient that dust particles can retain charge following ionisation. During photoionisation the electrical conductivity of particles is observed to vary by up to 6 orders of magnitude but tends to return to its original value once the ionisation source is removed (Carrier et al. 1991). The magnitude of this conductivity change is observed to be highly dependent on the temperature at which photoionisation occurs and the energy of the incident ionising photons. The net charge of lunar dust on the day side as a result of photoionisation is therefore expected to be positive . The near surface environment will be characterised by a photoelectron layer boundary between the surface and the incident solar wind. On the night side of the Moon charging of lunar dust particles occurs as a result of thermal currents of subsonic electrons and supersonic flows of ions and dust particles are likely to become negatively charged (Horanyi, 1996). The balancing potential for the surface dust in this later case may be as high as ~ 1 kV (e.g. Halekas et al., 2005). Local differences in topography, and the resultant illumination and composition may also have a significant effect on the actual charging and potential of dust. A global scale potential difference may therefore exist between the day and night sides of the Moon across the terminator (Stubbs et al., 2007c) and across topographical boundaries between light and dark. Measurements from Lunar Prospector demonstrate a typical potential difference between the illuminated and non-illuminated sides of the Moon of around 100V (Halekas et al., 2011). Electric fields across the terminator may become very large when solar wind fluxes are high and a plasma wake is formed at the night side (Stubbs et al, 2007c). While traversing through the Earth's magnetosphere the Moon's surface is subject to a plasma environment which is significantly different from that of the solar wind, being much more tenuous but of much higher energy. Under these circumstances the conditions for surface charging will be changed considerably (Halekas et al., 2011; Collier et al., 2011). . The integrated lunar environment at a south polar landing site is illustrated in Figure 3.

An additional source of dust arises through a combination of micrometeoroids, which impact at hypervelocity and the ejecta particles generated by these impacts on the lunar surface. These particles will also result in a continuous, low density population of impactors, incident on the surface (Grün et al., 2011). Thus three separate dust populations are present; levitated dust, ejecta and meteoroids. The fluxes and relative contributions of each of these populations to the overall incident dust flux at the surface is not known and is at present poorly constrained. In measuring these populations they may distinguished through a combination of their velocities, trajectories and charges.

It is also important to consider that, regardless of the effects on dust particles and the interactions between those particles and systems, the charge and plasma environment its-self and its interaction with surfaces is also of high importance for future exploration activities. Charging of lander surfaces and the potential for large local potential differences (and thus perhaps fields) on scales comparable with surface operations, must also be system drivers for future exploration missions. Again however the minimal quantity of data available, particularly from the surface, leaves the true nature of the environment unknown and generates a requirement for in situ surface measurements.

The limited quantity of in situ measurements of the lunar dust environment made to date have led to significant uncertainties regarding any models of such dust and the implications for future exploration. As such new in situ measurements of the key parameters characterising such populations and those which drive the important and likely physical processes are required. An objective for the mission is therefore to improve current models of charging, transport, adhesion and abrasion of lunar dust as relevant to future lunar exploration activities. Some of the key parameters, for which uncertainties exist, for which further measurements are required, and which are considered for the Lunar Lander mission are discussed below.

*The size distribution of dust particles of the order for 10 nm and larger*. The particles with the greatest mobility, whose adhesion generates the greatest problems, whose behaviour is most difficult to predict and which have the greatest toxic potential are those with the smallest sizes and lunar dust particles can exist on scales of several nanometers and larger. For toxicity studies dust particles below around 10μm are considered respirable (although some authors consider lower values)and this constitutes around 10% of the lunar soil (the fraction with particles <1cm). Despite the importance of this size regime it is also the most incompletely characterised. This is in part because of the challenge associated with measurement, particularly during the epoch of collection, and in part because the geological and geochemical information of interest is available to a greater extent in the larger size fractions. A key parameter is the size distribution function, particularly in the sub-micron size regime. Recent attempts to measure the size distribution for the smallest particles (Park et al., 2008) have led to questions of the extent to which samples in the existing collection can be considered representative of in situ dust when the smallest size fraction (<1μm) is considered. Such uncertainty arises because the collection and curation of these particles was not performed with a consideration for this smallest particles and it is likely that the size distribution function has been biased, during the chain from collection, to curation, to investigation. A review of lunar dust and its properties is provided by Liu and Taylor (2011).

*The extent of nanophase Fe in lunar dust*. A unique property of lunar dust is the presence of nanophase metallic iron particles, whose sizes can vary between ~3 nm and 30 nm. These particles are generated by reduction of Fe bearing minerals during micrometeoroid impacts exposure to high energy particles. In the smallest size fractions the abundance of Fe nanoparticles is the greatest (Taylor et al. 2001). These particles are generally encased in a glassy matrix however impact glass may be easily dissolved in bodily fluids, releasing the nanophase Fe grains, which are likely to be very reactive, owing to their redox potential and large surface area per unit volume. It may also be the case that some of these particle are found at the surfaces of the glassy matrix and may thus be directly exposed to tissue. In this case toxicity may be significantly enhanced. Thus the presence, extent and surface distribution of nanophase Fe is of high importance in determining the toxicity of particles. In addition nanophase Fe, being both conductive and paramagnetic, may also affect the behaviour of dust particles in the presence of ionising radiation, plasma, charging and electric and magnetic fields. Thus characterising nanophase Fe is also important for the understanding of the behaviour of mobile dust and dust surface interactions.

*The mineralogical composition of dust grains*. While the mineralogy of dust grains in sample collections is fairly well understood for previously explored sites, that expected at the lunar south pole remains to be analysed. Mineralogy of particles is an important consideration for the determination of the toxicity of particles. Variations in mineralogy may also affect key physical properties (e.g. dielectric constant, work function etc.) underlying their physical behaviour. Mineralogical measurements of the larger size fractions In the regolith can also be used to identify its potential as a feedstock for resource extraction processes (see later section). It should also be noted that the mineralogy of regolith at the lunar south pole will include that of minerals excavated from the South Pole-Aitken Basin, potentially exposing upper crust and lower mantel material, considered of high importance for planetary science investigations (e.g. Crawford et al., 2012; this issue, Committee on the Scientific Context for Exploration of the Moon, 2007).

*The structures of Lunar dust grains < 10µm with a goal of extending this to grains > 10nm*, The structures of grains are of potential importance for studies of toxicity and can also give insight into formation mechanisms, electrostatic properties and bulk mechanical properties. A study of the structures of the smallest grains in existing samples was performed by Liu et al. (2008). Investigations on the Lunar Lander can provide the properties of lunar dust grains from a new and un-sampled particle population, from an area of high relevance for future missions and access the properties of those grains as they exist in-situ.

*The elemental composition of dust grains*. The elemental composition of dust grains is an important addition to mineralogical measurements and can help constrain particle composition. It can be particularly important in the case of glassy materials for which their amorphous nature prevents measurement by some techniques (e.g. Raman spectroscopy) which targeting mineralogy. Elemental composition provides an important input for toxicity studies, and in the larger size fractions provides information relevant to resource identification and geochemistry.

*The adhesion of dust grains as a function of solar incidence angle and material properties.* As discussed earlier the adhesion of particles to surfaces appears to be related to solar illumination and thus observations to confirm and quantify this are important. In addition the extent of adhesion to materials with differing properties is also important for the definition of future exploration systems. This can also provide insight in to the interactions between dust particles and human tissues.

*Charges on levitating dust grains*. The charges on grains are related to the surface potential and the composition, structure and morphology of individual grains. Under some simple assumptions though it can be shown that typical charges on sub-micron grains may be in the range 0.1 – 10 fC. Thus a requirement for measurements should be to measure charges on levitated grains of 1fC and less with a goal of achieving 0.1 fC.

*Velocities of mobile dust grains.* The velocities of dust grains levitated electrostatically from the surface must vary significantly as a function of particle size, with the smallest particle having the greatest velocities. The LADEE mission to be launched in 2013 will carry a dust instrument (Horanyi et al., 2009) sensitive to dust particles on the 0.5µm scale at altitudes as low as 30km. Such altitudes correspond to particles with vertical velocities at the surface of greater than 300m/s, assuming that such particle move on ballistic trajectories and neglecting electrostatic effects after ejection. In order to provide synergy with measurements made by LADEE therefore the upper limit for determination of velocity of ejected particles should be at least 300m/s, with a goal sensitivity to particles with higher velocities. Being closer to the surface however the Lunar Lander will be subject to particles which may be larger than those observed in orbit and which have lower velocities. Thus a sensitivity to the slowest particle possible is preferred. A nominal slowest particle velocity of 1m/s might be expected for a mobile particle.

*Trajectories of levitating dust particles.* Mobile dust particles are comprised of three populations, levitated dust from the surface, ejecta on ballistic trajectories and meteoroids from space. The primary mechanism for discrimination between these populations is the direction of incidence. A measurement of incident direction to an accuracy of 5° is expected to be sufficient to allow discrimination. Measurements should include both vertical and horizontal components of particle velocity.

*Flux of mobile particles*. A key measurement for defining the effects of any mobile dust population on future exploration systems is the particle flux of the various dust populations and the variation of this value as a function of the environmental conditions. At present there are very few constraints on these values.

*Electric fields.* The acceleration of levitated dust particles must be accomplished via electric fields through which particles travel. Ideally both vertical and horizontal components of the electric field should be measured. A priority is most likely on vertical electric field as this is the component which must initiate levitation. Typical electric fields are expected to be in the range ±10 V/m (Stubbs, 2007c; Sternovsky et al., 2008).

*The density and temperature of the local photoelectron layer.* During periods of illumination the surface near surface plasma environment will be dominated by "cold" photoelectrons, with energies typically around 1eV. The extent of this sheath is expected to be within a few Debye lengths of the surface, though measurements of this would be of benefit. The number density of these particles will be highly dependent on the angle of incidence of solar radiation at the give surface but it typically expected to be around $10^3$ cm$^{-3}$ following measurements made during Apollo (Reasoner and Burke 1972; Feuerbacher et al. 1972; Walbridge, 1973) and maybe as high as $10^4$ cm$^{-3}$. At the lunar poles this is expected to be lower given the typically reduced angle of incidence and may be closer to $10^{-2}$ cm$^{-3}$. Electric fields with this electron sheath are important for the mobilisation of dust particles and the temperature and density of this layer are also important parameters for understanding the dynamics of this charged particle population and the interaction between it, the surface and incident plasma populations from the solar wind and magnetospheric plasma populations.

*The electron and ion energies in incident solar wind and magnetospheric plasmas.* Both solar wind and magnetosphere plasmas are incident on the lunar surface. Solar wind plasmas are incident for the majority of the lunar orbit, while for around 25% of the lunar orbit the Moon passes through the Earth's magnetosphere and is subject to magnetospheic plasma populations. A review of the lunar plasma environment is provided by Halekas et al. (2011). These incident populations are also vital elements in the integrated dust-radiation-plasma-surface, environment. In addition chemical interactions between incident solar wind and surface materials are likely to provide a source of water on the Moon, as discussed in the following sections. Thus it is important to characterise the properties of these incident populations and the relationship between them, surface charging, the local plasma environment and the charging, behaviour and mobility of dust particles. Incident particles will also result in the sputtering of surface materials, yielding a low density plasma population of surface sputtered ion and secondary electrons. Solar wind particles are typically considered to be "cold" with energies as low as a few eV and number densities of around 10cm$^{-3}$ (e.g. Zombeck, 2007). Magnetospheric particles in the tail lobes however are far less abundant, having number densities of around 0.1 cm$^{-3}$ (Escoubet et al., 2007) but have much greater energies as high as tens of keV. The directionality of these population is also important. While magnetospheric plasma may be considered to be incident isotropically on local scales the solar wind is highly directional and measurement of the distribution of the angular and velocity distributions for particles is also important.

## 3.3 Radiation environment and effects

The surface of the Moon is subject to various types of ionising radiation, which constitute a very significant hazard for human operations. Radiation is dominated by four distinct populations: low energy Solar Wind Particles (SWPs), high energy Galactic Cosmic Rays (GCRs), sporadic high energy particles released during Solar Energetic Proton (SEP) events and secondary radiation generated by the interactions of these primary sources with the lunar surface and subsurface to depths of approximately a meter. Secondary radiation will also be generated by particle interactions with materials in spacecraft and lunar surface infrastructure. In addition ionisation of the surface, to depths of several microns, results from solar UV and X-ray photons. Understanding and then mitigating this environment is essential for the future human exploration of the Moon. A comprehensive discussion of the lunar radiation environment is provided by Reitz et al. (2012; this issue). Lunar astronauts will be exposed to these various sources of radiation. It is known that protons and high-atomic-number energetic particles (HZE) may have significant biological effects, even at low fluences, and considerable uncertainties exist about the effects of secondary particles. Indeed the long term (chronic) and short term (acute) effects of radiation may be the primary limiting factor for the potential of human beings to explore beyond low Earth orbit. A comprehensive discussion of the current understanding and recent work into the effects of radiation expected for deep space explorers is provided by Durante and Cuccinota (2011) and

Durante (2012; this issue). It is clear that major uncertainties exist with regard to the risks and effects of HZE particles in human physiology but investigations into the biological processes involved are complex. Investigations have been performed to explore and potential ways to address some of these uncertainties from a small in situ precursor mission using in situ microscopy of sub cellular DNA damage by HZE particles and subsequent repair (Carpenter et al. 2010; de Vos et al., 2012; this issue) but the timescales associated with development of such experiments are unlikely to be compatible with the Lunar Lander mission. Thus for the Lunar Lander mission the objective is to monitor the incident radiation and received dose.

### 3.4   Identifying resources for future exploration

In Situ Resource Utilisation (ISRU) is the term used to refer to the generation of consumables for autonomous or human activities from raw materials found in-situ. The use of ISRU in the future may provide a means of reducing the ultimate cost and risk of operation on the Moon and provide a means for commercial contributions to lunar exploration. Potential products include $O_2$ and $H_2O$ for life support or $H_2$ and $O_2$ for fuel and propellant (also potentially by hydrazine production from $N_2$, $NH_3$ and $H_2O_2$). The most comprehensive discussion on the potential mechanisms for extraction of resources from Lunar regolith is currently given in "The Resources of Near Earth Space" (Ed. Lewis et al. 1993). Recent results however, particularly the confirmation of the presence of water in polar regions of the Moon, have extended the possibilities and likelihood of in situ resources being utilised significantly (see Crawford et al., 2012; this issue and Anand et al., 2012; this issue).

A sustainable future for lunar exploration is likely to depend upon the effective use of in situ resources to generate products such as oxygen, water and other used consumables. The choice of which ISRU processes might be applied in the future is dependent on both technical maturity and the availability and distribution of resources that might be used. At present the status of both of these elements is considered to be insufficient to allow the selection of a single ISRU process for demonstration at the lunar surface at a given landing site, particularly one as different from those explored by Apollo as the lunar polar regions. It is anticipated that the First Lunar Lander can best contribute to future ISRU activities by aiding the determination of availability, distribution and abundance of potential in-situ resources, in order to inform the future development of ISRU capabilities and mission planning. The mission will seek to improve our understanding of the potential use of polar regolith as a resource. It has been demonstrated from measurements of returned samples, lunar meteorites and remote sensing data that lunar crustal materials and regolith composition are diverse across the lunar surface and that measurements of lunar regolith and rocks, as well as sample return missions have provided analysis of material from sites which are known to be unrepresentative of the whole lunar surface. Regolith and crustal material in near polar regions in particular may be quite different from that observed in current sample collections (see for example Crawford et al. 2012; this issue and references within). Knowledge of lunar highlands material comes from analyses of remote sensing data and lunar meteorites, and a limited number of Apollo samples. Future missions to lunar highlands and non-mare locations (e.g. the lunar South Pole) at which ISRU is to be applied will require a process which is appropriate to the available feedstock, in this case lunar highlands regolith. To this end it is beneficial to determine the properties of this feedstock in terms of its mineralogical, chemical and elemental composition, and physical properties and behaviour during handling when in situ on the Moon.

*Identify the presence or otherwise of water ice outside of permanently shadowed craters.* It has long been suggested that water ice might be present in permanently dark and near polar craters on the Moon (e.g. Arnold, 1979). The interior of these craters have been shown to be as cold as 29 K by the Diviner instrument on LRO (Paige et al., 2010). Increased levels of hydrogen at both the North and South lunar poles were confirmed by Lunar Prospector (up to ~1700 ppm) (Feldman et al. 1998 and 2000). Recent measurements by the L-CROSS mission (Colaprete et al., 2010) have confirmed that the water is indeed present. While of tremendous interest from a future exploration perspective the topographic location of this water and the challenging environment in which it resides may prove prohibitive for excavation for some time yet. For the Lunar Lander this water is definitely not accessible. However indications of subsurface temperature from Diviner (Page et al. 2010) and epithermal neutron fluxes from the Lunar Exploration Lunar Detector (LEND) (Mitrofanov et al., 2010) instruments on LRO, have raised the possibility that, in the near polar regions outside of permanently shadowed areas, the sub-surface, at depths of less than a meter, may also contain significant quantities of water ice. Identifying the presence or otherwise of water ice in the subsurface at a known site in the polar regions is an objective for the lunar lander.  It should be noted however that the intended landing site for the Lunar Lander will be at a local topographical high point, with extended periods of illumination. Such sites are unusual and, while important for their exploration potential may also, by virtue of the extended illumination periods, not be locations where subsurface water may be found. As such a negative observation cannot be inferred to extend further to the wider polar region. For this reason it is important that measurements are not limited to water, but are extended to a wider range of volatile chemistry.

*The abundance as a function of depth and distribution of $H_2O$, OH and hydrated minerals in the regolith.* Results from the missions Chandrayaan, Deep Impact and Cassini (Pieters et al, 2009; Sunshine et al., 2009, Clarke, 2009) have revealed the presence of either hydroxyl (OH) or water (H2O) or both in the lunar regolith through identification of spectral absorption features near to 3μm in wavelength. This feature corresponds to the fundamental vibration of absorption for the OH chemical group. Initial indications are that up to a few tenths of a percent by weight of water might be present to optical depths of a ~ 1 mm at the surface. The abundance of water observed appears to increase as a function of latitude so that the highest concentrations are observed close to the poles.

Variations in absorption have also been noted for different geological features, in particular associated with plagioclase feldspar (Pieters et al., 2009). Time variations in the strength of the absorption feature indicate that the water is dynamic and migrating across the lunar surface, possibly providing a source for ice trapped at the lunar poles although the validity of inferred temporal changes are not yet confirmed.

The presence of water at the lunar surface has potential implications for ISRU in future lunar exploration both as a mechanism for the generation of water as a resource in its self and also as a source for oxygen and hydrogen for fuel, through electrolysis. If water is present in significant quantities in the lunar soil then its extraction as a resource might be feasible. In order to determine if this is the case the distribution of water as a function of depth into the surface should be determined, as must its origins. The lunar Lander will target such measurements. However it should also be noted that alteration of the surface local to the landing site during the vehicle's descent is likely to make surface measurements unreliable. Instead measurements of volatiles contained within the subsurface provide perhaps the best opportunity for observations of water.

*The abundance and distribution of solar wind implanted volatiles in a non-Apollo locality.* Solar wind implanted volatiles are another potential resource, with the advantage over others that their extraction from the regolith can be achieved by heating alone. A discussion of the utilization of volatiles is provided by Fegley and Swindle (1993). These volatiles have a number of potential applications, including their use as a feed product for the reduction of element oxides (e.g. FeO) for the production of O and $H_2O$. Solar wind implanted volatiles include the elements H, N, C and He. These elements are important for various aspects of the maintenance of a lunar outpost and so volatile extraction may reduce the requirements for replenishment of these elements from a terrestrial source.

For all volatile populations the mission should seek to achieve the following:

- liberate volatiles and/or their volatile precursors from lunar regolith samples from known locations
- determine the chemical identities of volatile species or, if relevant their precursors within the lunar regolith
- provide a quantitative measure of the total volatile yield and the yields for individual volatiles as a function of the sample size. Where appropriate quantifying any precursor within the regolith which produces a volatile during the extraction process
- determine the isotopic abundance of the relevant volatile species as a differential measure relative to the accepted international standards. Such measurements can reveal the sources and processes that have lead to the emplacement of volatiles. Such an understanding is required to infer general trends on the distributions of volatiles based on measurements made at a single locality.

## 3.5 Preparing for scientific utilisation of the Moon

A significant part of the exploration efforts of humans living and working on the Moon will be directed at performing fundamental research (Battrick et al., 1992, Jaumann et al, 2012; this issue, Crawford et al, 2012; this issue). Investigations will be in diverse areas, from the origin and history of the Moon, to using the Moon as an enabling platform for astronomy, physical sciences and life sciences research. As a contribution to ensuring the sustainability of lunar exploration Europe's first Lunar Lander can help optimise output for the early missions and aim to perform science that might be unavailable during later stages of exploration.

Future human missions to explore the Moon are likely to engage in science investigations from the Moon, which address otherwise inaccessible science goals. In the case of astronomy the Moon's stable surface, large solid volume and exposure to almost free-space may play an important role, following some of the most remote and deserted areas on Earth. On Earth some of the regions most hostile to life that have become a focus of major research activities in geophysics, climate research, astronomy and astrophysics. For example, one hundred years after the exploration of Antarctica, its ice is now being used as a cubic kilometre detector of ultra-high energy neutrinos in the IceCube project, and the extremely high-altitude deserts and mountains of the Andes host some of the biggest optical and radio telescopes in the world, with a host of smaller experiments following suite.

It is conceivable that within a couple of decades the Moon will offer a similarly popular focus for a wide range of science investigations and a goal of recent exploration studies has been to identify the most promising applications for science on and from the Moon. An important area, for which the Moon may be enabling, is low-frequency radio astronomy and astrophysics (e.g. Klein-Wolt et al., 2012; this issue). This is because the far side of the Moon is expected to offer a uniquely radio-quiet environment, outside the Earth's ionosphere, that allows the study the earliest phase of the observable universe and planetary plasma environments.

An additional consideration, for missions coming in advance of a surge in surface exploration activities is the Moon's Surface Boundary Exosphere (SBE). This is a tenuous atmosphere produced by vaporisation of surface material, by ion sputtering and meteoritic impacts, under conditions where collisions between individual atoms and molecules are extremely rare and liberated gases may therefore have long parabolic trajectories back to the surface or may escape from the surface all together (Mendillo, 2001). Other bodies in the solar system which have SBEs include Mercury, Europa, Ganymede, Callisto and Enceladus and SBEs may be present on other satellites and Kuiper belt objects. SBEs are poorly understood and the Moon offers the best, and perhaps only, opportunity to perform detailed studies of their sources and dynamics. Currently the composition and properties of the Moon's SBE are poorly known and significant investigation is required to improve knowledge and extend the understanding gained to other bodies in the Solar System. The total mass of the lunar exosphere is of the order of 100 tonnes. This compares with a mass of exhaust gasses ejected during landing of several tonnes. As a result once human exploration begins in earnest the environment will quickly become dominated by the gaseous products of these activities and measurements of the exosphere will no longer be possible. Thus early exploration missions will provide the only opportunity to study the natural lunar exosphere and plasma environment.

The Scientific Context for Exploration of the Moon (Committee for the Scientific Context for Exploration of the Moon, 2007) was generated in 2007 by the National Research Council in the USA. It presents a community consensus of the various topics and objectives for scientific investigation on the Moon and identifies those objectives of highest priority. The highest priority objectives presented are accepted here as correct. Note that objectives related to utilisation of the Lunar surface as a platform for observations (e.g. for astronomy and astrophysics) are not included in this list. It is assumed that research performed on the surface of the Moon in the future will in large part be driven by these top priority objectives, which are as follows:

1. Test the cataclysm hypothesis by determining the spacing in time of the creation of the lunar basins.
2. Anchor the early Earth-Moon impact flux curve by determining the age of the oldest lunar basin (SPA).
3. Establish a precise absolute chronology.
4. *Determine the composition and distribution of the volatile component in lunar polar regions.*
5. Determine the extent and composition of the primary feldspathic crust, KREEP layer, and other products of planetary differentiation.
6. Determine the thickness of the lunar crust (upper and lower) and characterize its lateral variability on regional and global scales.
7. Characterize the chemical/physical stratification in the mantle, particularly the nature of the putative 500-km discontinuity and the composition of the lower mantle.
8. *Determine the global density, composition, and time variability of the fragile lunar atmosphere before it is perturbed by further human activity.*
9. Determine the size, composition, and state (solid/liquid) of the core of the Moon.
10. *Inventory the variety, age, distribution, and origin of lunar rock types.*
11. *Determine the size, charge, and spatial distribution of electrostatically transported dust grains and assess their likely effects on lunar exploration and lunar-based astronomy.*

While the Lunar Lander mission has not been defined with a view to addressing fundamental scientific questions it should be noted that, in addressing the exploration enabling science questions described in the above sections, the mission will in any case contribute to science priority numbers 4, 8, 10, 11 and 12 in the above list. As such the mission will make a significant contribution to fundamental scientific questions and begins the process of using exploration as a means of allowing scientific exploitation of the Moon.

# 4 Model Scientific Payload

While at the time of writing a final identification of the mission's payload has not been performed, a model payload has been derived for application in the mission study and to inform on going scientific activities. The model payload does not represent a final selection of the payload. The model payload has been derived by taking the objectives and requirements described, identifying potential experimental techniques that can be applied to address these areas of research, considering the level of maturity of the relevant technologies and then combining these with the main mission boundary conditions as defined. The intersection of the science objectives and requirements, the boundary conditions of the mission and the maturity and requirements of possible experiments defines those objectives which may feasibly be achieved and potential experiments for the mission.

The model payload is an essential element in the design and development of the mission in advance of formal payload selection. The model payload informs the mission study of likely payload requirements and interfaces to allow the system to be designed appropriately, identifying potential challenges and problem areas relating to specific payload types.The model payload can be used to identify areas where developments in payload design, instrumentation and associated technologies are required at payload or lander system level. Finally the investigation of the model payload provides insight into the data sets that might be expected from the mission and allows scientific preparatory work to be performed with a view to maximising the mission's scientific return.

Of importance in the derivation of a model payload of the mission was the recognition that the primary objective for the mission is the demonstration of soft precision landing with hazard avoidance. Thus it is this requirement which drives the mission design. The interfaces and requirements associated with any payload considered for the mission must be compatible with a mission derived in this context. Payloads whose requirements adversely affect the realisation of the primary objective cannot be considered as this may jeopardise the implementation of the mission. It is vital to ensure that the payload selected remains compatible with the nominal mission and its boundary conditions while maintaining a high scientific return. The boundary conditions to be considered for the payload are summarised below:

*Environment.* The payload must be compatible with the environmental conditions implied by a landing at the lunar poles. This includes a challenging thermal environment.

*Duration and operations cycle.* A landing at the lunar pole is intended to maximise the mission duration possible without the application of radioisotopes for thermal control or power. To this end a mission duration of no longer than 6 to 8 months is expected. A minimum possible mission of nominally one month should be considered. Payload investigations must be compatible with this operational duration.

*Mass.* The mission will launch on a Soyuz from Kourou. This results in an upper limit on the total mass of the mission, including the lander platform and the payload. Once the platform has been sized appropriately and robustly to ensure that the primary objective is achieved, remaining mass is available for payload and the required support infrastructure (e.g. robotic arm, sampling system, batteries, harness, attachments etc.). The total mass available for payload and related items is under investigation and cannot be confirmed at the time of writing but is not expected to exceed ~40 kg.

*Lander System.* The Lunar Lander system will be defined to optimise the technology demonstration aspects as this is the mission's primary objective. Payload related requirements will be accommodated by the system design in as much as they do not compromise the primary objective and do not incur significant cost and risk penalties, thus jeopardising the robustness and feasibility of the mission. The model payload considered during the studies is presented in Table 1. The table also refers to the measurements that will be made and the investigation topics that are addressed by these measurements.

In some cases it has been noted that a number of instruments have requirements, objectives, interfaces and operational characteristics that lend themselves to an increased level of integration. In these cases the experiments are grouped together into packages. In these cases further consideration of the scientific complementarity, operational synergy, co-location and integration have been considered. This approach seeks to optimise the science return from an instrument whilst optimising the use of resources and reducing the complexity of payload to system interfaces. At the time of writing these experiment packages are the subject of on-going definition and design studies to ensure feasibility for the mission and prepare for a future call for experiments. These studies are described further below. It is important to note that the measurements performed by each instrument, or package cannot be considered in isolation. A full utilisation of the scientific data set requires that the payload and the resultant data be considered as a whole, and by an interdisciplinary science community. Considering single instruments, or even packages of instruments, in isolation will result in a significant reduction in scientific return from the mission and will not achieve the objective of addressing the key scientific problems associated with exploration of the moon.

The experiment packages identified in Table 1 are under currently under study. These payload packages are referred to as the Lunar Camera Package (L-CAM); the Lunar Dust Analysis Package (L-DAP); the Lunar Dust Environment and Plasma Package (L-DEPP); the Lunar Volatile Resources Analysis Package (L-VRAP); Radiation monitor.

## 4.1 Lunar Dust Environment and Plasma Package (L-DEPP)

The Lunar Dust Environment and Plasma Package (L-DEPP) (e.g. Hausmann et al., 2012; this issue) is a package whose function is to determine the charging, levitation and transport properties of lunar dust, in-situ on the Moon, and the associated properties of the local plasma environment and electric fields. As described the lunar dust has been identified as the primary potential hazard for operations on the surface of the Moon and for the development of future lunar surface infrastructure. The properties of lunar dust are however very poorly understood. Particular uncertainty surrounds the charging, levitation and transport of lunar dust particles in the integrated dusty plasma of the lunar environment. Other properties associated with the local plasma and electromagnetic environment may have additional implications for exploration systems design and development.

The scientific requirements addressed by the L-DEPP suite of instruments is discussed in Table 1. In summary the package aims to measure the temperature and density of the local plasma the magnitude of local electric fields and observe the radio spectrum as an additional means of probing plasma properties (with an additional goal to prepare for future radio astronomy activities).

Possible configurations for instruments in the package are shown in Figure 5 and Figure 6. The package includes a dust particle charge and trajectory sensor operating on a principle of charge induction in wire grids (Srama et al., 2007) to measure the charges on levitating lunar dust particles and their velocities and trajectories. Electric fields and the temperature and density of cold plasma, including the local electron sheath, are measured by Langmuir probes (for heritage see e.g. Holback et al. 2001 and Eriksson et al. 2007). The optimal configuration for these probes is yet to be determined. In order to measure all three components of the local electric field four probes are required on booms, however mass limitations and mission level accommodation constraints are likely to limit the number of probes which can be deployed meaning that priority must be given to components of the field to be measured. The energies and directions of ions and electrons are determined through the application of an ion / electron spectrometer.

The addition of a radio antenna to the package can provide vertical profiling of the lunar ionosphere and plasma properties by tracking known sources across the sky. In addition characterising the long wavelength radio environment at the landing site is an essential step towards the eventual realisation of a lunar radio observatory.

### 4.2 Radiation monitor

The nominal radiation monitor is a derivative of the IRAS instrument (Federico et al., 2007) which was originally intended for ESA's Exomars mission. The radiation monitor provides dosimetry and ground truth for models of the lunar radiation environment. The verification of such models is a key to their successful application in the planning and implementation of future exploration missions, in particular with regard to determining the effects for human explorers. For a further description of the Radiation monitor see Reitz et al. (2012; this issue).

### 4.3 Lunar Dust Analysis Package (L-DAP)

The Lunar Dust Analysis Package is a package for the in-situ microscopy and compositional analysis of lunar dust and regolith. As described in earlier sections the lunar regolith, and in particular the smallest size fraction, is both a hazard for lunar exploration and a potential asset. While potentially toxic to humans and hazardous to surface infrastructure dust also contains potential resources which, if extracted, may be key to the sustainability of future exploration. The microscopic properties of lunar dust are integral to understanding all of these factors and a suite of instruments operating in synergy on a lunar lander can provide information required when developing future lunar surface infrastructure. The science requirements addressed by the package are shown in Table 1. The experiment is an integrated package of instruments comprising an optical microscope, atomic force microscope and Raman spectrometer for making observations of particles which are sampled and delivered to the package by the lander's sampling system / robotic arm, who's reach is expected to be approximately 2m. The inclusion of magnetic force microscopy to the atomic force microscope's capabilities may also allow investigation of nanophase iron in grains and its distribution (Staufer et al., 2012; this issue). Figure 4 illustrates a likely configuration for the L-DAP package.  In addition an external Raman spectrometer and Laser Induced Breakdown spectrometer optical head, accommodated on the mission's robotic arm allows the measurement  of mineralogy and elemental composition for objects outside of the lander. This allows the characterisation of surfaces in advance of samples being taken and measurement of objects which are too large to be sampled and observed on board. This external head works in conjunction with a camera, also located on the robotic arm. The highly integrated nature of the package ensures optimisation of available resources from the lander and promotes a highly synergistic approach to science operations and data acquisition

The instruments in the package build on significant experiment heritage obtained through previous missions. The atomic force microscope, sample stage and optical microscope have been integrated as a package as part of the MECCA experiment, developed for NASA's Phoenix mission. (Hecht et al., 2008). MECCA was originally defined as a human exploration precursor experiment. The optical microscope also draws on the heritage of the microscope developed for Beagle 2 (Thomas et al. 2004). A Raman instrument is currently considered as part of the payload of the Exomars mission and the integrated Raman-LIBS spectrometer considered here is derived from the elegant breadboard developed in the frame of preparations for Exomars (Courreges-Lacoste et al., 2007).

## 4.4 Lunar Volatile Resources Analysis Package (L-VRAP)

The Lunar Volatile Resources Analysis Package (L-VRAP) (Sheridan et al., 2012; this issue) measures the species of volatiles present at the lunar surface, their abundance and distribution from a landed platform. The package also demonstrates their extraction, as a precursor to future in-situ resource utilisation in human missions. Recent observations have shown that extensive quantities of volatile elements and molecules are found near to the poles of the Moon. These volatiles are potentially the most important resources on the Moon, enabling sustainable long term exploration of the Moon and potentially fuel for exploration beyond the Moon. In planning for future surface infrastructure and human exploration activities it is important to understand and characterise the availability and distribution of volatiles in the lunar regolith. The experiment package is required to extract the volatile molecules from lunar soil samples delivered to it by a robotic sampling system. The samples are then heated incrementally in ovens to temperatures to temperatures as high as 1200°C. The gasses liberated during this heating are then analysed by a mass sector mass spectrometer to determine the species present, their isotopes and relative abundance.

It is clear that measurements made of volatiles in the lunar regolith in the vicinity of the lander will be subject to contamination and alteration. These effects will primarily result from the interactions of the thruster plumes, with the surface during the descent. This will both implant the surface with volatile species resulting from the combustion of the fuel (monomethyl hydrazine and mixed oxides of nitrogen) and cause physical abrasion of surface dust and regolth layers. Because the potential effects of contamination by the Lander will be critical to measurements, current activities continue to investigate the extent of these affects, their important and methods for mitigating these effects. For mitigation three approaches can be considered:

1) Sampling from sufficient depths beneath the surface that contamination is not observed.
2) Sampling from large lateral distances from the lander
3) Characterising the chemistry of contaminants such that they can be identified and removed from measurements.

For case 1) preliminary results indicate that contamination may be limited to the upper few centimetres of the surface. If this were demonstrated to be the case in future work then , then sampling of the subsurface, even at shallow depths, may be sufficient to remove the majority of contamination. For this reason current studies are investigating a system to take regolith samples, in the close vicinity of the lander and from depths of greater than 10 cm and nominally up to 40cm.

For case 2) a translational capability is required on scales, likely to be several hundreds of meters. Accommodating such a capability as part of a small payload is expected to be challenging and the feasibility of doing so cannot be assured. For this reason the feasibility of collecting samples from approximately 100m from the lander and returning them via a small robotic element has also been investigated as a means of collecting samples with minimal contamination and alteration (Haarmann et al 2012, this issue).

For case 3) the capabilities of the L-VRAP instrument are employed to identify and remove contamination on the basis of the chemical and isotopic signatures of contaminants. A goal for future work therefore will be to characterise the specific chemical signatures of contaminats and to then demonstrate that these can be recognised and removed these from measurements. A further discussion to approaches to mitigating contamination in L-VRAP measurements is provided by Sheridan et al. (2012; this issue).

The system applied in the model payload is derived from the Gas Analysis Package on Beagle 2 (Wright et al., 2000) and the Ptolemy Instrument on Rosetta (Todd et al., 2007). The configuration of the L-VRAP package is shown in Figure 8.

The same instrument package is also under investigation as a means of measuring the species in the lunar exosphere. It this case the exosphere particles for examination enter the package via a passive atmospheric particle collector, which is open to space. Once collected atmospheric particles are analysed using an ion trap mass spectrometer.

## 4.5   Lunar Camera Package (L-CAM)

The baseline Panoramic Camera uses wide angle stereo imaging and narrow angle monoscopic high resolution (colour) imaging to acquire information on it surroundings, while multiple narrow band filters allow e.g. mineral composition of rocks and soils to be measured. The nominal camera system is derived from the PanCam system for Exomars (Griffiths et al., 2006, Coates et al., 2012; this issue), though some adaptation may be required for the lunar environment. The addition of close a imaging camera on the lander's robotic arm allows for close up analysis of the surface, of areas where samples may be taken, of objects to be analysed spectroscopically of the lander its self, including robotic operations.

## 4.6   Surface Operations and Sampling Strategy

Once landed on the surface of the Moon the Lunar Lander will deploy its high gain antenna and make contact with ground. This will be followed by the deployment of L-CAM, which will image soundings and the horizon to establish the illumination profile for the mission and verify the location of the landing. Deployment of L-DEPP instruments and any other instrument requiring deployment would follow and monitoring of the environment by L-DEPP and L-RAD would begin. L-CAM panoramic images would be generated at regular intervals throughout the mission. Of particular interest for L-DEPP are the properties of the environment across light-dark boundaries. To this end the lander shall endeavour to provide power and data resources to L-DEPP as the Lander moves from illumination into darkness and vice versa.

Prospective areas of interest for measurements or sampling shall first be identified and located using images from L-CAM, which provides a context for measurements to be made or samples to be taken. The areas of interest shall then be approached using the lander's robotic arm. Close up images from the L-RAC shall then be used to provide a more detailed context for the areas of interest. If required then mineralogical and chemical measurements can be made using the L-DAP Raman/LIBS external optical head. For rocks and objects too large to be sampled and introduced to the lander's internal analysis suite the LIBS laser can be used to ablate surfaces and provide access to internal material for which space weathering is reduced.

Where an area is deemed to be of interest for sampling the mission's sampling device shall then be deployed to the surface. The device can then enter the surface and take samples from depths from the very surface to 10cm. A goal shall be to target samples form depths of up to 40cm. Samples taken at depth are then transferred by the robotic arm to the L-VRAP and / or L-DAP instrument packages for analysis. A challenge for the sampling system is to minimise alteration of the samples during the extraction and delivery process. Samples are then deposited into the payload of choice. The payload its self is then responsible for any further processing or manipulation of the samples and distribution to the various instruments in the packages.

# 5    Conclusions and future work

The ESA Lunar Lander mission offers an opportunity to prepare the way for future exploration activities of the Moon and beyond through the development of new technologies and the generation of scientific knowledge that enables the exploration programmes of the future. Key knowledge gaps associated with future exploration have been identified, for which in situ measurements are required. These knowledge gaps cover a wide array of scientific topics in a variety of scientific disciplines. In addressing scientific objectives related to exploration preparation it is also possible to make measurements of relevance to a number of high priority fundamental scientific objectives identified for lunar exploration.

A model payload has then been defined, which offers the possibility to address these unknowns as far as is possible within the mission constraints.  These constraints are defined by the physical limitations of the mission and it's launcher, the requirement to use a power system using solar arrays and batteries only and the requirements imposed by the primary objective of the mission which is to validate soft safe precision landing  technologies.  The experiments in this payload address the integrated plasma-radiation-charge-dust environment and its effects, the properties and effects of lunar dust, the chemical and mineralogical resources available at the lunar south pole including water and other volatiles, and the topography, geology and physical properties of an important landing site for future exploration missions.

# 6 References

1. Anand, M., Crawford, I., Balat-Pichelin, M., Abanades, S., van Westrenen, W., Péraudeau, G., Jaumann, R., Sebolt, W., 2012, A brief review of chemical and mineralogical resources on the Moon and their potential utilization, This issue
2. Arnold, J.R. , 1979, Ice at the lunar poles, *J.Geophys.Res.,* 84, 5659–5668.
3. Battrick, B., Barron, C. and the Lunar Study Steering Group, 1992, Mission to the Moon. Europe's priorities for the scientific exploration and utilisation of the Moon, European Space Agency, Paris (France), Jun 1992.
4. Berg, O.E., et al. , 1976, Lunar soil movement registered by the Apollo 17 cosmic dust experiment, in Interplanetary Dust and Zodiacal Light, Springer-Verlag, Berlin, 233–237.
5. Bergmann, J., 2012, Electric field measurements for investigating the dynamics of dusty plasma on the lunar surface, this issue
6. Bonnefoy, R., Link, D., Casani, J., Vorontsov, V.A., Engström, F., Wolf, P., Jude, R., Patti, B., Jones, C. , 2004, The Beagle 2 Commission of Enquiry, http://www.bis.gov.uk/assets/bispartners/ukspaceagency/docs/space%20science/beagle-2-commission-of-inquiry-report.pdf.
7. Bussey et al., 2012; this issue
8. Carey et al., 2012, The Moon as a stepping stone to Mars: the human robotic partnership, This issue
9. Carpenter, J. D., Houdou, B., Koschny, D., Crawford, I., Falcke, H., Kempf, S., Lognonne, P., Ricci, C., Pradier, A., 2008, The MoonNEXT Mission: A European Lander at the Lunar South Pole, A. Joint Annual Meeting of LEAG-ICEUM-SRR, October 28-31, 2008 in Cape Canaveral, Florida. 1446, 33.
10. Carpenter, J. D., Angerer, O., Durante, M., Linnarson, D., Pike, W. T. , 2010, Life Sciences Investigations for ESA's First Lunar Lander, *Earth, Moon, and Planets*, 107, 1, 11-23.
11. Carrier, III,D., Olhoef, and Mendell, W., 1991, Physical properties of the Lunar surfaces, in Lunar Sourcebook, Ed. Heiken, G.H. Vaniman, D. and French, B.M., Cambridge University Press.
12. Carroll, W.F. and Blair, Jr., P.M. , 1972, Lunar dust and radiation darkening of Surveyor 3 surfaces, analysis of Surveyor 3 material and photographs returned by Apollo 12, NASA SP-284, 158-167.
13. Chin, G., Brylow, S., Foote, M., Garvin, J., Kasper, J., Keller, J., Litvak, M., Mitrofanov, I., Paige, D., Raney, K., Robinson, M., Sanin, A., Smith, D., Spence, H., Spudis, P., Stern, A.S., Zuber, M., 2007, Lunar Reconnaissance Orbiter Overview: The Instrument Suite and Mission, *Space Science Reviews*, 129, 4, 391-419

14. Clark, R.N. , 2009, Detection of adsorbed water and hydroxyl on the Moon, *Science*, 326, 5952, 562.
15. Coates, A., Griffiths, A., Leff, C., Schmitz, N., Barnes, D., Josset, J-C., Hancock, B., Cousins, C., Jaumann, R., Paar, G., Bauer, A., Crawford, I., 2012, Lunar PanCam: adapting ExoMars PanCam for the ESA Lunar Lander, This issue.
16. Cockell, C.S., 2010, Astrobiology - What can we do on the Moon?, *Earth Moon and Planets*, 107, 3-10.
17. Colaprete, A., Schultz, P., Heldmann, J., Wooden, D., Shirley, M., Ennico, K., Hermalyn, B., Marshall, W., Ricco, A., Elphic, R.C., Goldstein, D., Summy, D., Bart, G.D., Asphaug, E., Korycansky, D., Landis, D., Sollitt, 2010, L., Detection of Water in the LCROSS Ejecta Plume, *Science*, 330, 6003, 463.
18. Collier, M.R., Kent Hills, H., Stubbs, T.J., Halekas, J.S., Delory, G.T., Espley, J., Farrell, W.M., Freeman, J.W.,  Vondrak, 2011, R., Lunar surface electric potential changes associated with traversals through the Earth's foreshock, *Planetary and Space Science*, 59, 14, 1727-1743.
19. Committee on the Scientific Context for Exploration of the Moon, 2007, the Scientific Context for Exploration of the Moon, National Academies Press.
20. Courreges-Lacoste, B, Ahlers, B., Rull, F. ,  2007, Combined Raman spectrometer/laser-induced breakdown spectrometer for the next ESA mission to Mars, *Spectrochimica Acta Part A: Molecular and Biomolecular Spectroscopy*, 68, 4, 1023-1028.
21. Crawford, I.A., 2004, The scientific case for renewed human activities on the Moon, *Space Policy*, 20, 91-97, 2004.
22. Crawford, I. A., Anand, M. ,  Cockelle, C. S.,  Falcke, H., Green, D. A., Jaumann, R., Wieczorek, M. A., 2012, Back to the Moon: The Scientific Rationale for Resuming Lunar Surface Exploration, this issue
23. De Rosa et al., 2012; this issue
24. De Vos, W., Meesen, G., Szpirer, C., Scohy, S., Assouak, S., Evrard, O., Hutsebaut, X., Beghuin, D., 2012, Real time detectionof radiation effects in individual cells on the surface of the Moon, a cytomics strategy for Lunar biodosimetry, This issue.
25. Durante, M., 2012, Space radiobiology on the Moon, This issue

26. Eriksson, A. I., Boström, R., Gill, R. , 2007, RPC-LAP: The Rosetta Langmuir Probe Instrument, *Space Sci. Rev*., 128, 1-4,. 729-744.

27. Escoubet, C. P., Pedersen, A., Schmidt, R. and Lindqvist, P. A., 1997, Density in the magnetosphere inferred from ISEE 1 spacecraft potential, J. Geophys. Res., 102, 17, 595-609

28. Feldman, W.C., Maurice, S., Binder, A.B., Barraclough, B.L., Elphic, R.C., Lawrence, D.J., 1998. Fluxes of fast and epithermal neutrons from Lunar Prospector: evidence for water ice at the lunar poles, *Science*, 281, 1496–1500.

29. Feldman, W.C., Lawrence, D.J., Elphic, R.C., Vaniman, D.T., Thomsen, D.R., Barraclough, B.L., Binder,A.B. , 2000, Chemical information content of lunar thermal and epithermal neutrons, *J. Geophys. Res*. 105, 20347–20363.

30. Federico, C., Di Lellis, A. M., Fonte, S. et al. , 2007, EXOMARS IRAS (DOSE) radiation measurements, Memorie della Società Astronomica Italiana Supplement, 11, 178.

31. Fegley, B., Jr. and Swindle, T. D., 1993, Lunar volatiles: implications for lunar resource utilization, Resources of near-Earth space, Ed. Guerrieri, M.L., University of Arizona Press, Arizona, USA. 367 - 426

32. Feuerbacher, B., Anderegg, M., Fitton, B., Laude, L. D., Willis, R. F., & Grard, R. J. L. , 1972, Photoemission from lunar surface fines and the lunar photoelectron sheath, Proc. Lunar Sci. Conf. 3rd, 2655.

33. Fisackerly, R., Pradier, A., Philippe, C., Houdou, B.. De Rosa, D., Carpenter, J., Gardini, B., 2011, ESA Lunar Lander Mission, proc. 8th International Conference on Guidance Navigation and Control, Karlovy Vary, Czech Republic, June 2011.

34. Klein Wolt, M., Aminaei, A., Zarka, P., Schrader, J-R., Boonstra, A-J., Falcke, H., 2012, Radio astronomy with the Lunar Lander: opening up the last unexplored frequency regime, This issue.

35. Glenar, D.A., Stubbs, T.J., McCoy, J.E., Vondrak, R.R., 2011, A reanalysis of the Apollo light scattering observations, and implications for lunar exospheric dust, Planetary and Space Science, 59, 14, 1695–1707.

36. Goswami et al., 2010, Using the Moon as an analogue to study biobehaviour, this issue

37. Griffiths, A.D., Coates, A.J., Jaumann, R., et al. , 2006, the Context for the ESA ExoMars rover: the Panoramic Camera (PanCam) instrument. *International Journal of Astrobiology*, 5, 3. 269-275.

38. Grün, E., Horanyi, M., Sternovsky, Z., 2011, The lunar dust environment, *Planetary and Space Science*, 59, 14, 1672-1680.

39. The global exploration strategy: a framework for coorperation, , 2007, http://esamultimedia.esa.int/docs/GES_Framework_final.pdf.
40. Haarmann, R., Mobile Payload Element (MPE): Concept Study for a Sample Fetching Rover for the ESA Lunar Lander Mission, This issue
41. Halekas, J. S., Lin, R. P., Mitchell, 2005, D. L., Large negative lunar surface potentials in sunlight and shadow, *Geophysical Research Letters*, 32, 9, CiteID L09102.
42. Halekas, J. S., Saito, Y., Delory, G. T., Farrell, W. M. , 2011, New views of the lunar plasma environment, *Planetary and Space Science*, 59, 14, 1681-1694.
43. Hausmann, G., Burfeindt, J., Bernhardt, H-G., Bergman, J., Falke, H., Klein-Wolt, M., Srama, R., Hofmann, p., Richter, L.,2012, Lunar Dust Environment and Plasma Package, this issue.
44. Hecht, M.H., Marshall, J., Pike, W.T., et al. , 2008, Microscopy capabilities of the Microscopy, Electrochemistry, and Conductivity Analyzer, *J.Geophys.Res.*, 113, E00A22, doi:10.1029/2008JE003077.
45. Holback, B., Jacksén, Å., Åhlén, L, et al. , 2001, LINDA - the Astrid-2 Langmuir probe instrument, *Ann. Geophys.*, 19, 601–610.
46. Horanyi, M. , 1996, Charged dust dynamics in the solar system, *Annu. Rev. Astron. Astrophys.*, 34, 383-418.
47. Horanyi, M., Sternovsky, Z., Gruen, E., Srama, R., Lankton, M., Gathright, D., 2009. The Lunar Dust EXperiment (LDEX) on the Lunar Atmosphere and Dust Environment Explorer (LADEE) Mission, Lunar and Planetary Institute Science Conference Abstracts, 40, 1741
48. Huixian, S., Shuwu, D., Jianfeng, Y., Ji, W., Jingshan, J., 2005, Scientific objectives and payloads of Chang'E-1 lunar satellite, *Journal of Earth System Science*, 114, 6, 789-794
49. Immer, C., Metzger, P. , Hintze, P., Nick, A. , Horan, R., 2010, Apollo 12 Lunar Module exhaust plume impingement on Lunar Surveyor III, *Icarus*, 211, 2, 1089-1102.
50. International Space Exploration Coordination Group, 2011, The global exploration roadmap, http://www.globalspaceexploration.org.
51. Jaumann R., et al., 2012, Lunar exploration and scientific results, this issue
52. Loftus, D.J., Rask, J.C., McCrossin, C.G., Tranfield, E.M., 2010, The chemical reactivity of lunar dust: from toxicity to astrobiology, Earth Moon and Planets, 107, 95-105
53. Katzan, C.M. and Edwards, J.L. , 1991, Lunar dust transport and potential interactions with power systems components, NASA contract NAS3-25266.

54. Lewis, J.S. , and Matthews, M.S., 1993, Resources of Near-Earth Space, Ed. Guerrieri, M.L., University of Arizona Press, Arizona, USA.

55. Linnarsson, D., Carpenter, J., Fubini, B., Gerde, P., Karlsson, L., Loftus, D., Prisk, K., Staufer, U., Tranfield, E., van Westrenen, W., 2012, Toxicity of lunar dust, This issue.

56. Liu, Y., Park, J., Schnare, D., Hill, E., Taylor, L.A., 2008, Characterization of lunar dust for toxicological studies. II: Morphology and physical characteristics, *J. Aerosp. Eng*. 21, 272, DOI:10.1061/(ASCE)0893-1321(2008)21:4(272).

57. Liu, Y., L.A. Taylor, 2011, Characterization of lunar dust and a synopsis of available lunar simulants, Planetary and Space Science, 59, 14, 1769-1783.

58. Lunar exploration definition team, 2009, Lunar exploration objectives and requirements definition document, ESA document LL-ESA-ORD-413, http://wsn.spaceflight.esa.int/docs/lunarlander/LunarLander_LERD_CDI_230512.pdf

59. McKay, D.S., Heiken, G., Basu, A, Blanford, G., Simon, S., Reedy, R., French, B.M. and Papike, J. , 1991, The Lunar Regolith, in Lunar Sourcebook, Ed. Heiken, G.H. Vaniman, D. and French, B.M, Cambridge University Press.

60. Mendell, W.W. and Heydorn, R.P., 2004, Lunar precursor missions for human exploration of Mars-III: studies of system reliability and maintenance, *Acta Astronautica*, 55, 773-780.

61. Mendillo, M., The atmosphere of the Moon, Proc. Earth-Moon Relationships, Padova, 2001.

62. Mitrofanov, I. G., Sanin, A. B., Boynton, W. V., Chin, G., Garvin, J. B., Golovin, D., Evans, L. G., Harshman, K., Kozyrev, A. S., Litvak, M. L., Malakhov, A., Mazarico, E., McClanahan, T., Milikh, G., Mokrousov, M., Nandikotkur, G., Neumann, G. A., Nuzhdin, I., Sagdeev, R., Shevchenko, V., Shvetsov, V., Smith, D. E., Starr, R., Tretyakov, V. I., Trombka, J., Usikov, D., Varenikov, A., Vostrukhin, A., Zuber, M. T. , 2010, Hydrogen Mapping of the Lunar South Pole Using the LRO Neutron Detector Experiment LEND, *Science*, 330, 6003, 483.

63. Murphy, T. W., Adelberger, E. G., Battat, J. B. R., Hoyle, C. D., McMillan, R. J., Michelsen, E. L., Samad, R. L., Stubbs, C. W., Swanson, H. E., 2010, Long-term degradation of optical devices on the Moon, *Icarus*, 208, 1, 31-35.

64. O'Brien, B. , 2008, Direct active measurements of movements of lunar dust: Rocket exhausts and natural effects contaminating and cleansing Apollo hardware on the Moon in 1969, *Geophys. Res. Lett*., 36, L09201, doi:10.1029/2008GL037116.

65. O'Brien, B. , 2011, Review of measurements of dust movements on the Moon during Apollo, *Planetary and Space Science*, 59, 1708–1726.

66. Paige, D.A., Siegler, M.A., Zhang, J.A., Hayne, P.O., Foote, E.J., Bennett, K.A., Vasavada, A.R., Greenhagen, B.T., Schofield, J.T., McCleese, D.J., Foote, M.C., DeJong, E., Bills, B.G., Hartford, W., Murray, B.C. Allen, C.C., Snook, K., Soderblom, L.A., Calcutt, S., Taylor, F.W., Bowles, N.E., Bandfield, J.L., Elphic, R., Ghent, R., Glotch, T.D., Wyatt, M.B., Lucey, P.G. , 2010, Diviner Lunar Radiometer Observations of Cold Traps in the Moon's South Polar Region, *Science*, 330, 6003, 479.
67. Park, J., Liu, Y., Kihm, K.D., Taylor, L.A. , 2008, Characterization of Lunar Dust for Toxicological Studies. I: Particle Size Distribution, *J. Aerosp. Eng. 21*, 266, DOI:1061(ASCE)0893-1321(2008)21:4(266).
68. Pieters, C. M., Goswami, J. N., Clark, R. N. et al. , 2009, Character and Spatial Distribution of OH/$H_2$O on the Surface of the Moon Seen by M3 on Chandrayaan-1, *Science*, 326, 5952, 568.
69. Pines, V., Zlatkowski, M., Chait, A., 2011 Lofted charged dust distribution above the Moon surface, Planetary and Space Science, 59, 14, 1795–1803.
70. Reasoner, D. L. and O'Brien, B. J. , 1972, Measurement on the lunar surface of impact-produced plasma clouds, *Journal of Geophysical Research*, 77, 1292 – 1299.
71. Reasoner, D. L., and W. J. Burke, 1972, Characteristics of the Lunar Photoelectron Layer in the Geomagnetic Tail, *J. Geophys. Res.*, 77, 6671
72. Reitz, G., Berger, T., Matthiae, D.l., 2012, Radiation Fields and Radiation Exposure at the Lunar Surface, This issue.
73. Sheridan, S., Wright, I.P., Morse, A.D., Merrifield, J.A., Waugh, L.J., Howe, C.J., Gibson, E.K., Pillinger., C., Barber., S., 2012, L-VRAP - a Lunar Volatile Resources Analysis Package for Lunar Exploration, This issue.
74. Srama, R., Srowig, A., Auer, S. , 2007, A Trajectory Sensor for Sub-micron Sized Dust, ESA SP-643: *Dust in Planetary Systems*, 643.
75. Staufer, U., Kohl, D., Schitter, G., 2012, The Potential of Magnetic Force Microscopy for in-situ Investigation of Nanophase Iron in Lunar Dust, This issue.
76. Sternovsky, Z., Chamberlin, P., Horanyi, M., Robertson, S. and Wang, X. , 2008, Variability of the lunar photoelectron sheath and dust mobility due to solar activity, *J. Geophys. Res.*, 113, A10104, doi:10.1029/2008JA013487.
77. Stubbs, T.J., Vondrak, R.R. and Farrell, W.M. , 2007a, Impact of Dust on Lunar exploration, in Dust in Planetary Systems, ESA SP-643.

78. Stubbs, T.J., Vondrak, R.R., Farrell, W.M. , 2007b, A dynamic fountain model for dust in the lunar exosphere, in *Dust in Planetary Systems*, ESA SP-643.
79. Stubbs, T.J., Halekas, J.S., Farrell, W.M., Vondrak, R.R. , 2007c, Lunar surface charging: a global perspective using Lunar Prospector data, in *Dust in Planetary Systems*, ESA SP-643.
80. Sunshine, J.M., Farnham, T.L., Feaga, L.M., Groussin, O., Merlin, F., Milliken, R. E., A'Hearn, M.F. , 2009, Temporal and spatial variability of lunar hydration as observed by the Deep Impact spacecraft, *Science*, 326, 5952, 565.
81. Taylor, L.A., Pieters, C.M., Keller, L .P., Morris, R. V. McKay, D.S. , 2001, Lunar mare soils: Space weathering and the major effects of surface-correlated nanophase Fe, *Journal of Geophysical Research*, 106, E11, 27985-28000.
82. Taylor, L.A., Schmitt, H.H., Carrier, W.D. III, 2006, The lunar dust problem: from liability to asset, proc. 1st Space Exploration Conference: Continuing the Voyage of Discovery, 30 January - 1 February 2005, Orlando, Florida.
83. Thomas, N.; Luthi, B.S.; Hviid, S.F.; Keller, H.U.; Markiewcz, W.J.; Blumchen, T.; Basilevsky, A.T.; Smith, P.H.; Tanner, R.; Oquest, C.; Reynolds, R.; Josset, J-.L.; Beauvivre, S.; Hofmann, B.; Ruffer, P. and Pillinger, C.T. , 2004, The microscope for Beagle 2, *Planetary and Space Science*, 52, 9, 853-866.
84. Todd, J., Barber, S., Wright, I., Morgan, G., et al, 2007. Ion trap mass spectrometry on a comet nucleus: the Ptolemy instrument and the Rosetta space mission, *Journal of Mass Spectrometry,* 42, 1, 1–10.
85. Wagner, S., 2006, The Apollo Experience Lessons Learned for Constellation Lunar Dust Management, NASA report, NASA/TP-2006-213726.
86. Walbridge, 1973, E., Lunar photoelectron layer, *Journal of Geophysical Research*, 78, 3668 – 3687.
87. Wright, I. P., Morgan, G. H., Praine, I. J., Morse, A. D., Leigh, D., Pillinger, C. T. , 2000, Beagle 2 and the Search for Organic Compounds on Mars Using GAP, proc. LPSC.
88. Zarka, P, Bougeret, J-L., Briand, C., Falke, H., Girard, J., Griessmeier, J-M., Hess, S., Klein-Wolt, M., Konovalenko, A., Lamy, L., Mimoun, D., 2012, Planetary and Exoplanetary Low Frequency Radio Observations from the Moon, This issue.
89. Zombeck, M.V., 2007, Handbook of Space Astronomy and Astrophysics, Cambridge University Press, Cambridge, UK.

# 7 Acknowledgements

The authors would like to acknowledge the extensive work performed by multiple groups in support of this work. These include the Lunar exploration definition team; the Topical Teams on Investigations in to the Biological Effects of Radiation (coordinated by Marco Durante, GSI), Toxicity of Lunar Dust (Coordinated by Dag Linnarson), Exploitation of local planetary materials (coordinated by Mahesh Anand, Open University), and Dusty Plasma Environments in the Solar System (coordinated by Jürgen Blum, Technical University of Braunschweig); The L-DAP Study team led by SEA Ltd.; the L-DEPP study teams led by the Finnish Meteorological Institute, Kayser Threde and The Czech Academy of Sciences; the L-VRAP study team led by the Open University; the Lunar Lander design team led by Astrium Space Transportation, colleagues throughout ESA who have supported this activity.

# Tables

Table 1. Science objectives and required measurements for the Lunar Lander model payload. Also shown are experiment packages and instruments within those packages which are applied to meet those requirements. Note that each objective requires data sets from multiple instruments. Thus extracting the optimal benefit from the payload requires that the payload and its various data sets be considered in a holistic and interdisciplinary way.

| Objective | Measurement | Instrument | Experiment package |
|---|---|---|---|
| **Site characterisation** | **Thermal environment** | **Temperature sensors** | **lander system** |
| | **Regolith mechanical properties** | **Robotic arm/sampling mechanism** | |
| | | **Cameras** | **L-CAM** |
| | **Illumination environment** | **Panoramic cameras** | |
| **Dust and plasma environment characterisation** | **Size distribution for particles > 10nm** | **Optical microscope** | **L-DAP** |
| | | **Atomic Force Microscope** | |
| | **Nanophase Fe in smallest grains** | | |
| | **Mineralogy of dust grains** | | |
| | **Structures of grains** | **Raman spectrometer** | |
| | **Elemental composition of grains** | **Laser Induced breakdown spectrometer** | |

| Objective | Measurement | Instrument | Experiment package |
|---|---|---|---|
| | Adhesion of grains | Sample wheel | |
| | Charges on levitated grains | Dust sensor | L-DEPP |
| | Velocities of mobile grains | | |
| | Trajectories of mobile grains | | |
| | Number density of mobile grains | | |
| | Electric fields | Langmuir probes | |
| | Density of photoelectrons | | |
| | Temperature of photoelectrons | | |
| | Vertical profile of photoelectron sheath | Radio antenna | |
| | Electron and ion fluxes | Electron and ion spectrometer | |
| | Electron and ion energies | | |
| | Electron and ion velocities | | |
| Radiation environment assessment | Incident solar particle fluxes | Radiation monitor | N/A |
| | Incident GCR fluxes | | |

| Objective | Measurement | Instrument | Experiment package |
|---|---|---|---|
| | Secondary particle fluxes | | |
| | Tissue equivalent dose | | |
| **Identifying and characterising resources** | mechanical properties | Robotic arm | lander system |
| | | Cameras | L-CAM |
| | Physical properties | Atomic Force Microscope | L-DAP |
| | | Thermal electrical permitivity probe | |
| | mineralogy of regolith | External Raman spectrometer head | |
| | elemental composition of regolith | Laser Induced breakdown spectrometer external head | |
| **Fundamental science precursors** | Radio Astronomy | Radio antenna | L-DEPP |
| | Exosphere | Mass spectrometer (ion trap) | L-VRAP |
| **Additional fundamental science achieved** | Exosphere | | |
| | Polar volatiles | Mass spectrometer (mass sector) | |
| | Inventory Rock types | Raman | L-DAP |

| Objective | Measurement | Instrument | Experiment package |
|---|---|---|---|
| | | Laser Induced breakdown spectrometer | |
| | | Cameras | L-CAM |
| | Electrostatic dust | Dust sensor | L-DEPP |
| | | Langmuir probes | |
| | | Electron and ion spectrometer | |
| | Radio astronomy preparation | Radio antenna | |

# Figures

Figure 1. Illustration of the descent and landing strategy for the Lunar Lander mission.

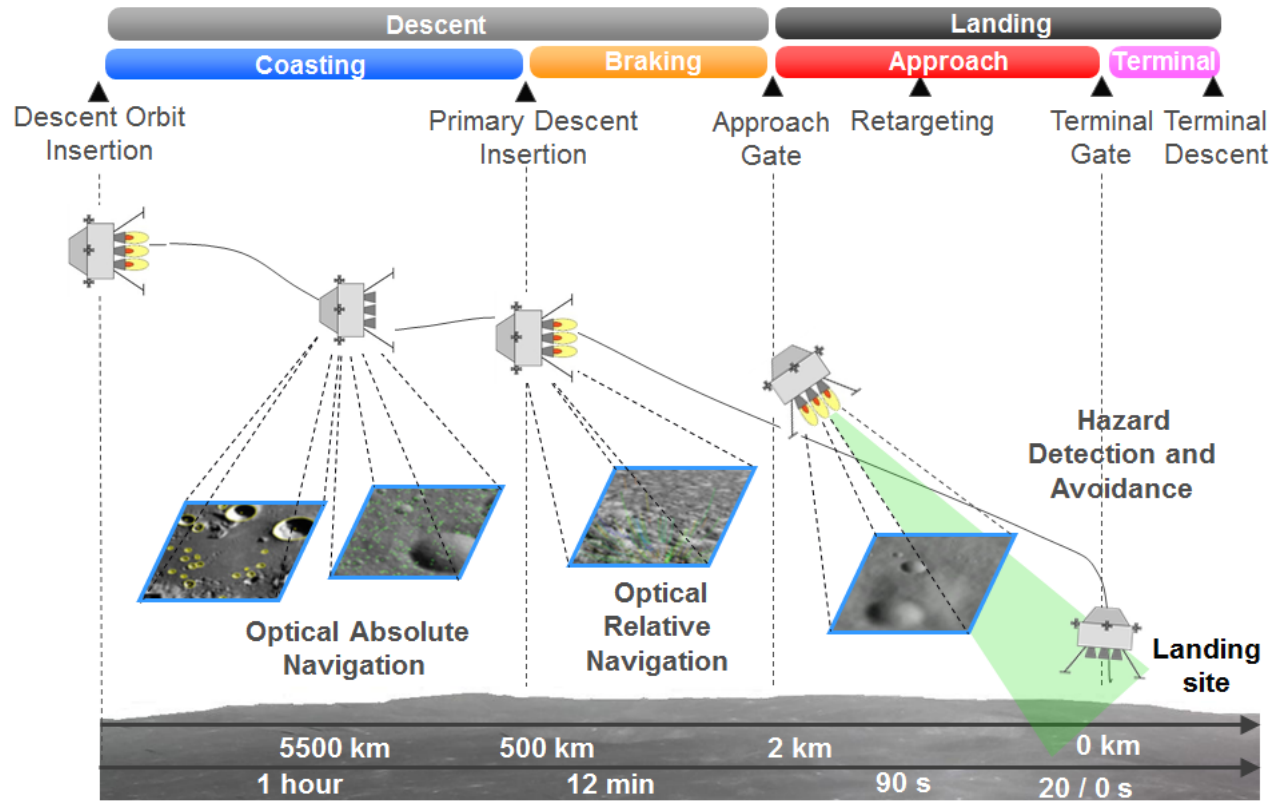

Figure 2. Current configuration of the Lunar Lander. Figure courtesy of Astrium.

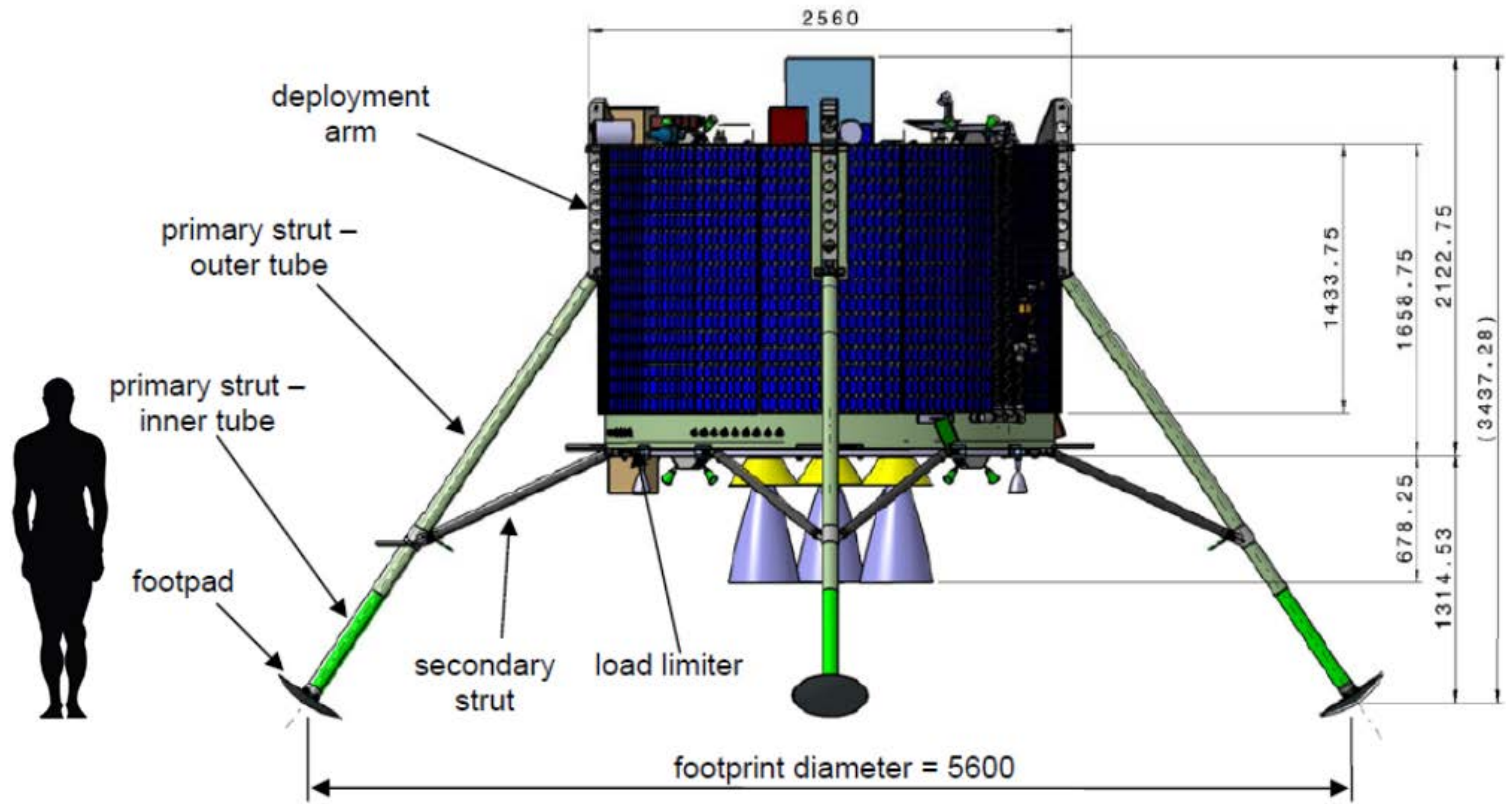

Figure 3. Illustration of a potential integrated environment to be measured at a lunar polar landing site. The surface in illuminated regions is charged positive by incident solar UV radiation. Just above the surface is a layer of photoelectrons. Thermal electrons in the solar wind can enter shadowed regions which charge negative. Other solar wind electrons are reflected away from the surface by the photoelectron cloud. Solar wind photons incident on the surface can reduce minerals forming $H_2O$/OH at grain surfaces. Magnetospheric plasma particles are incident when the Moon passes through the Earth magnetosphere. Charged dust particles can be accelerated through electric fields generated between the surface and the photoelectron sheath and between positive and negative charged surfaces.

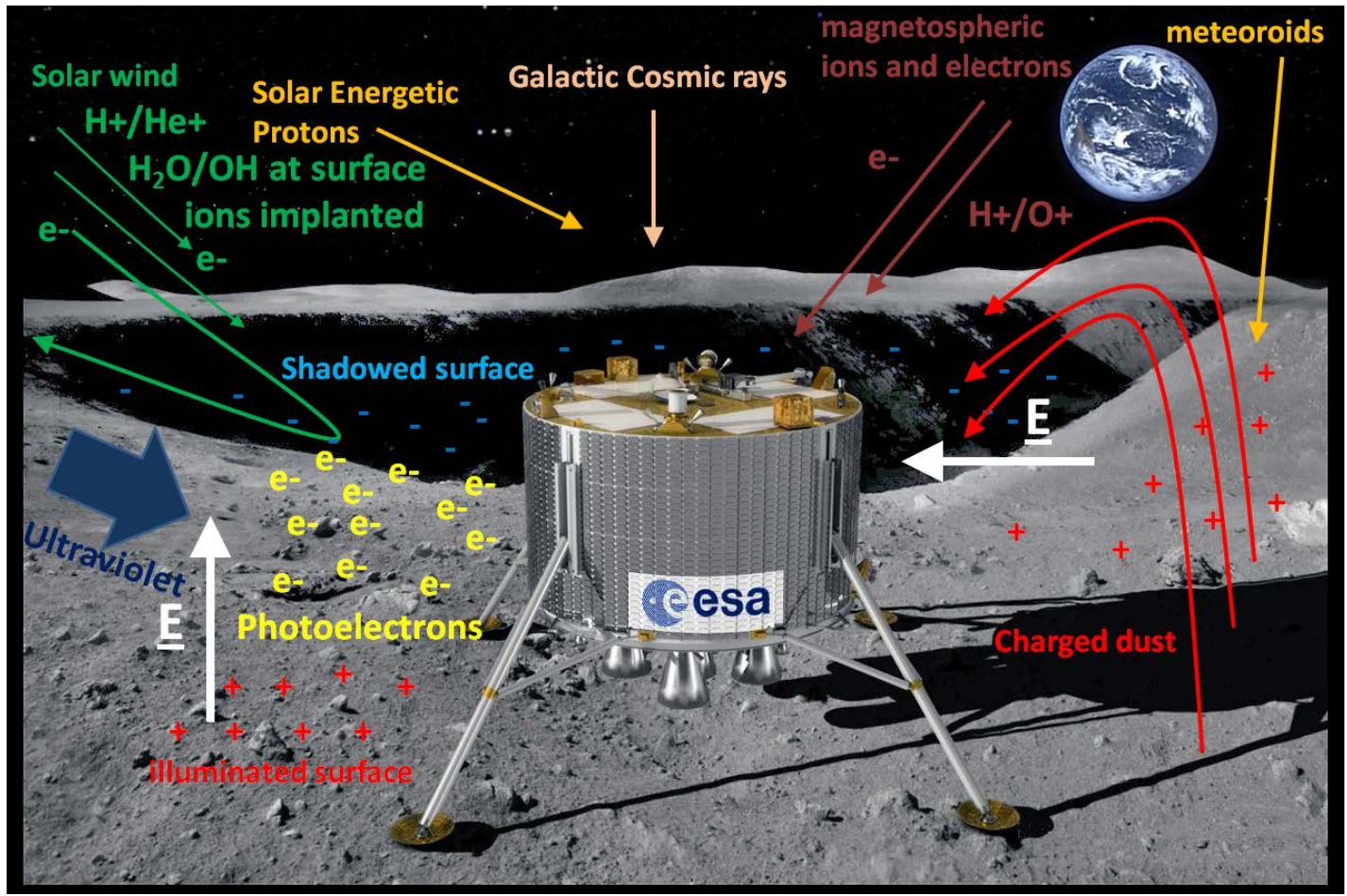

Figure 4. Nominal configuration of the Lunar Dust Analysis package. The optical microscope, atomic force microscope and Raman spectrometer internal head are mounted around a shared sample wheel allowing common samples to be analysed by all instruments in the package. The Raman/LIBS spectrometer unit is connected via optical fibres to an external optical head. Figure courtesy of SEA Ltd.

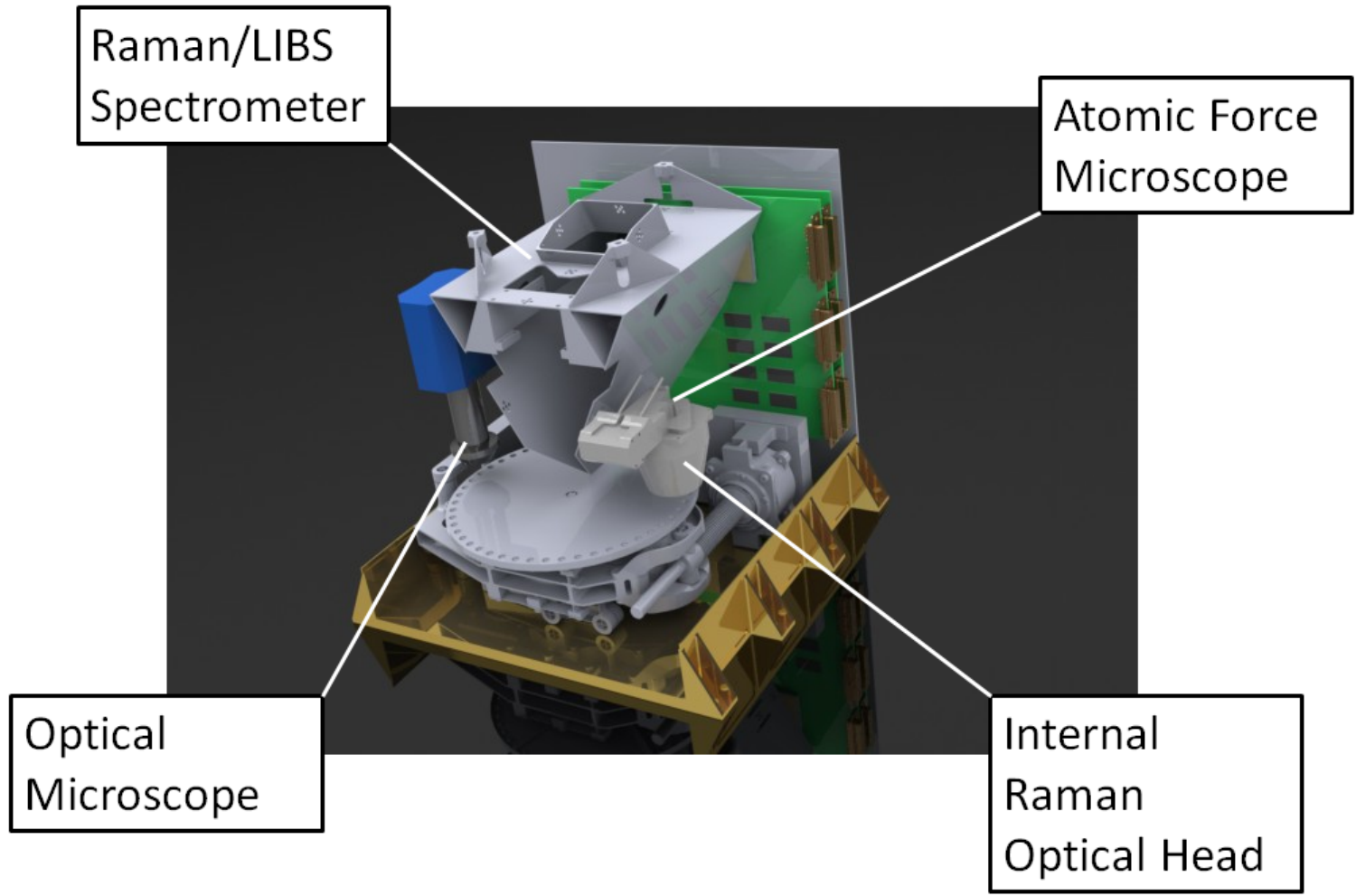

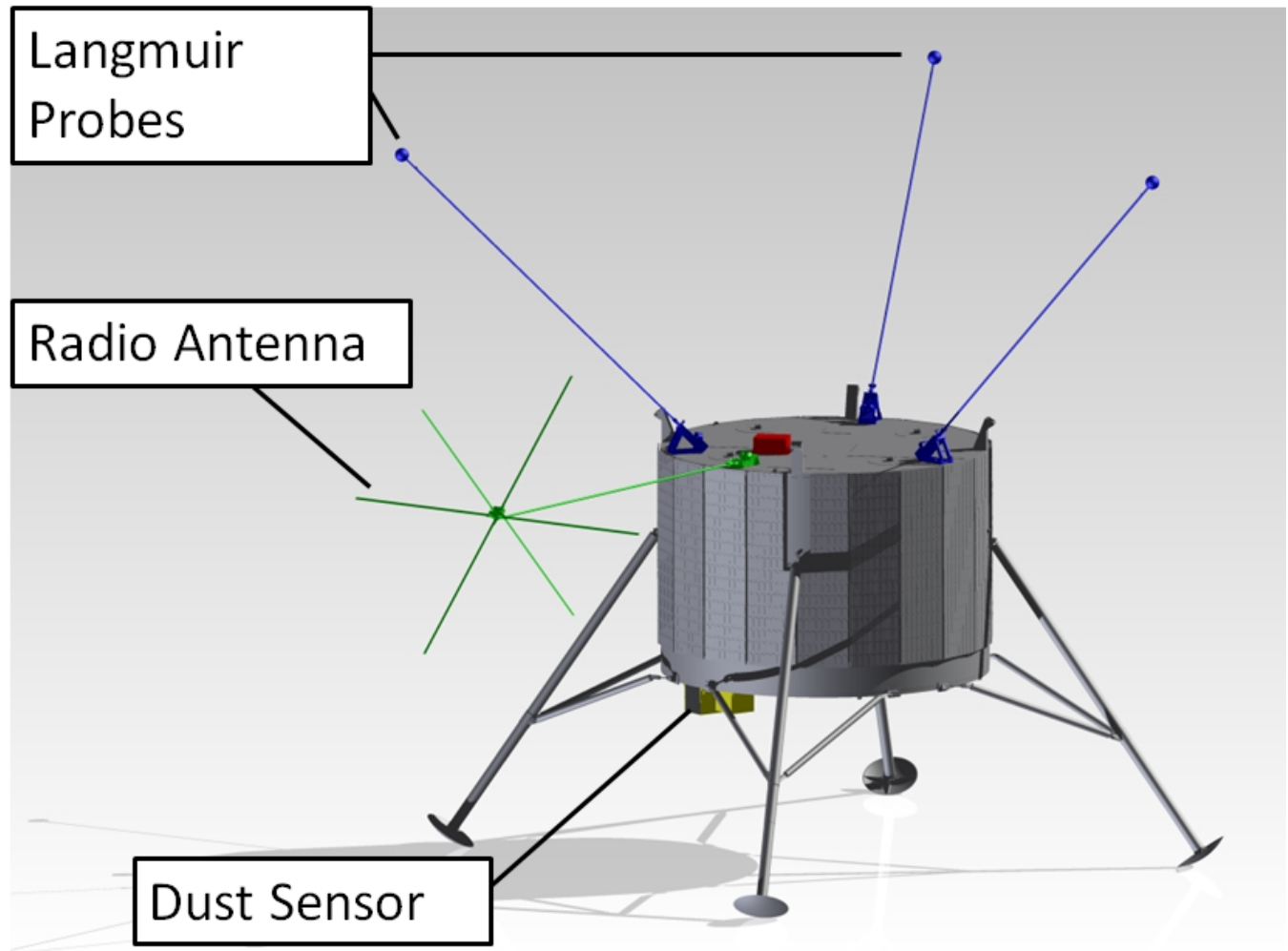
Figure 5. A potential configuration of the L-DEPP experiment package on the Lunar Lander. Figure courtesy of Kayser Threde.

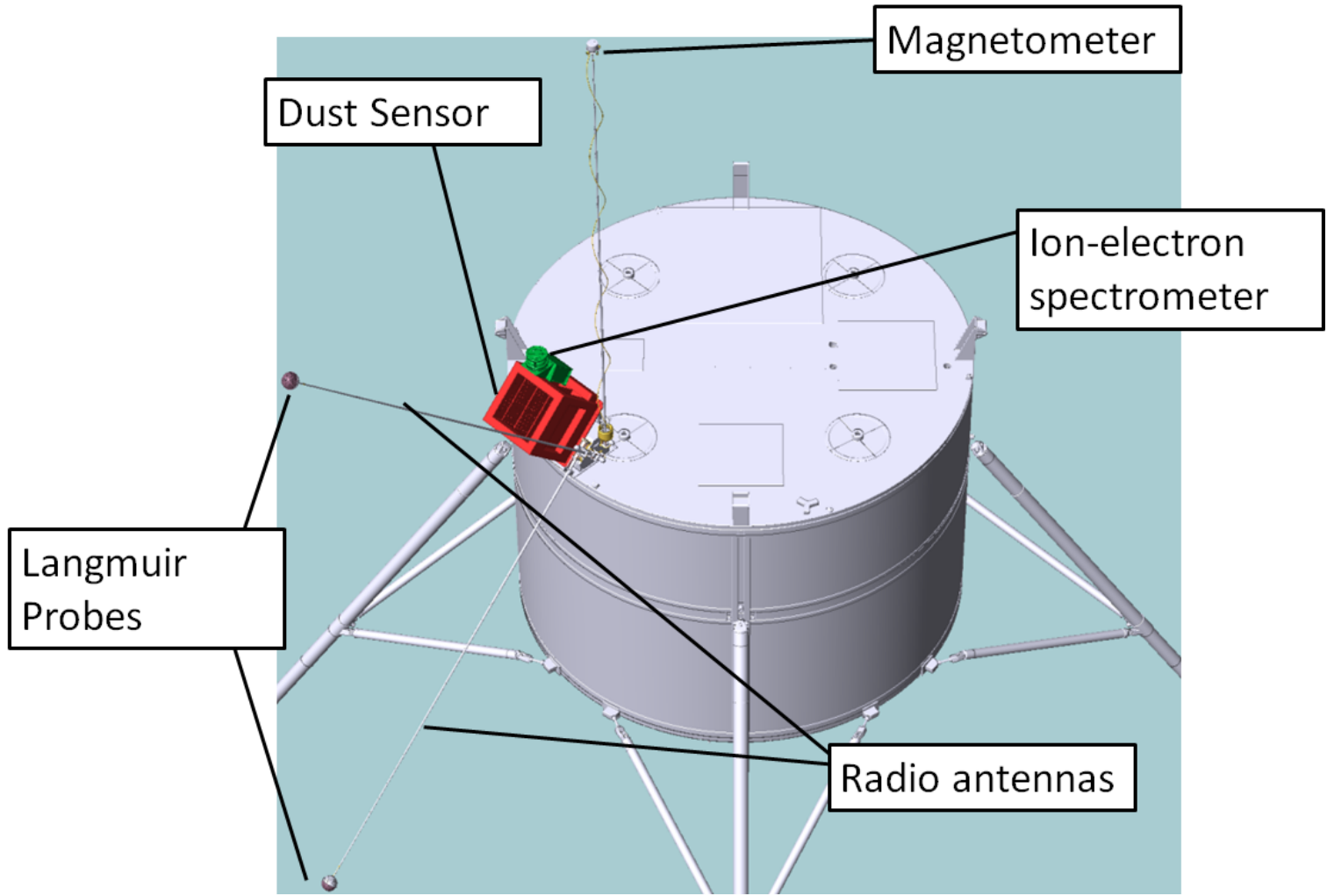

Figure 6. A potential configuration of the L-DEPP experiment package on the Lunar Lander. Figure courtesy of the Astronomical Institute of Czech academy of Sciences.

Figure 7. A potential configuration of the L-DEPP experiment package on the Lunar Lander. Figure courtesy of the Finnish Meteorological Institute and Swedish Institute for Space Physics.

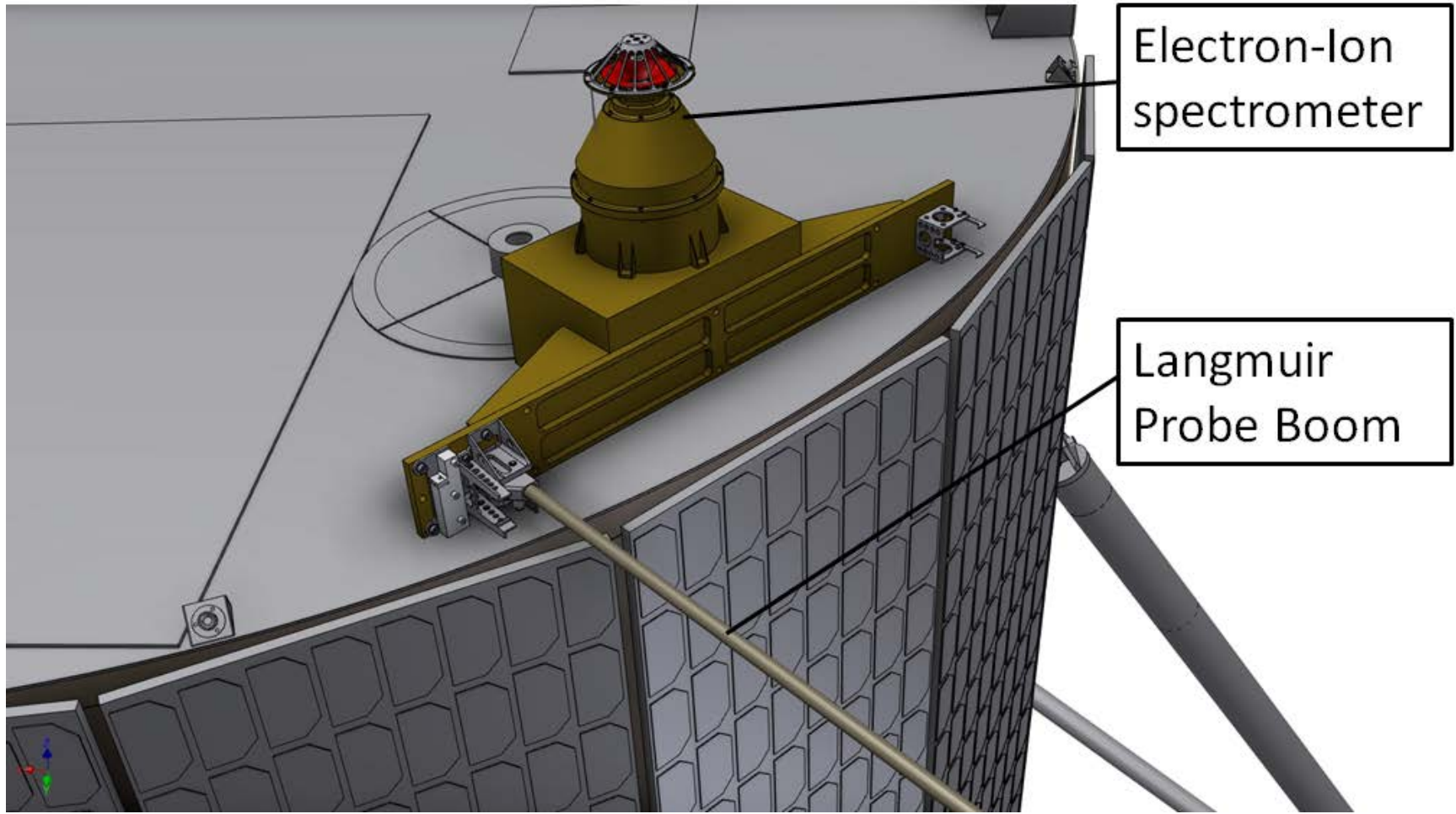

Figure 8. Early concept for the L-VRAP experiment package on the Lunar Lander. Figure courtesy of the Open University.

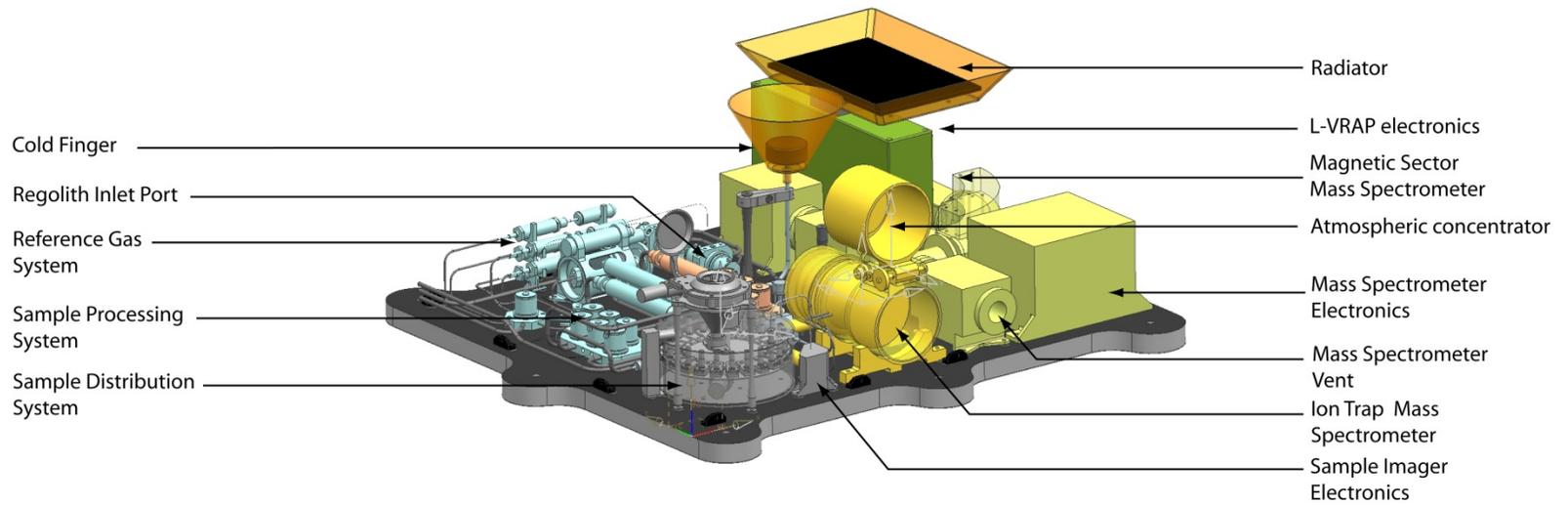